# Constraining Background N$_2$ Inventories on Directly Imaged Terrestrial Exoplanets to Rule Out O$_2$ False Positives

Sawyer Hall[1], Joshua Krissansen-Totton[1,2,3], Tyler Robinson[3,4,5,6], Arnaud Salvador[4,5,6], and Jonathan J. Fortney[1]
[1] Department of Astronomy & Astrophysics, UC Santa Cruz, Santa Cruz, CA 95064, USA; sajhall@ucsc.edu
[2] Department of Earth and Space Sciences, University of Washington, Seattle, WA, USA
[3] NASA Nexus for Exoplanet System Science Virtual Planetary Laboratory, University of Washington, Box 351580, Seattle, WA 98195, USA
[4] Lunar & Planetary Laboratory, University of Arizona, Tucson, AZ 85721, USA
[5] Department of Astronomy and Planetary Science, Northern Arizona University, Flagstaff, AZ 86011, USA
[6] Habitability, Atmospheres, and Biosignatures Laboratory, University of Arizona, Tucson, AZ 85721, USA
*Received 2023 April 26; revised 2023 September 30; accepted 2023 October 15; published 2023 November 20*

## Abstract

Direct imaging spectroscopy with future space-based telescopes will constrain terrestrial planet atmospheric composition and potentially detect biosignature gases. One promising indication of life is abundant atmospheric O$_2$. However, various non-biological processes could also lead to O$_2$ accumulation in the atmospheres of potentially habitable planets around Sun-like stars. In particular, the absence of non-condensible background gases such as N$_2$ could result in appreciable H escape and abiotic O$_2$ buildup, so identifying background atmosphere composition is crucial for contextualizing any O$_2$ detections. Here, we perform retrievals on simulated directly imaged terrestrial planets using `rfast`, a new exoplanet atmospheric retrieval suite with direct imaging analysis capabilities. By simulating Earth-analog retrievals for varied atmospheric compositions, cloud properties, and surface pressures, we determine what wavelength range, spectral resolution, and signal-to-noise ratio (S/N) are necessary to constrain background gases' identity and abundance. We find N$_2$ backgrounds can be uniquely identified with S/N ∼ 20 observations, provided that wavelength coverage extends beyond ∼1.6 $\mu$m to rule out CO-dominated atmospheres. Additionally, there is a low probability of O$_2$-dominated atmospheres due to an O$_2$–N$_2$ degeneracy that is only totally ruled out at S/N ∼ 40. If wavelength coverage is limited to 0.2–1.1 $\mu$m, then although all other cosmochemically plausible backgrounds can be readily excluded, N$_2$ and CO backgrounds cannot be distinguished. Overall, our simulated retrievals and associated integration time calculations suggest that near-infrared coverage to at least 1.6 $\mu$m and apertures approaching 8 m are needed to confidently rule out O$_2$ biosignature false positives within feasible integration times.

*Unified Astronomy Thesaurus concepts:* Exoplanet atmospheres (487); Spectroscopy (1558); Telescopes (1689); Astrobiology (74)

## 1. Introduction

With the Decadal Survey on Astronomy and Astrophysics 2020 prioritizing advances in astrobiology (National Academies of Sciences, Engineering, and Medicine 2021), and the HabEx and LUVOIR teams publishing their final report on their flagship mission concepts (The LUVOIR Team 2019; Gaudi et al. 2020), it is important to consider how future direct imaging telescopes could be optimized to distinguish atmospheric compositions and detect biosignature gases. One promising indication of life on Earth is abundant atmospheric O$_2$. However, various non-biological processes could also lead to O$_2$ accumulation in the atmospheres of potentially habitable planets around Sun-like stars (Kleinböhl et al. 2018; Meadows et al. 2018; Krissansen-Totton et al. 2021).

Planets in the habitable zones of M dwarf stars are potentially susceptible to abiotic oxygen buildup for several reasons. First, the elevated far-ultraviolet/near-ultraviolet ratios of their host stars may lead to the accumulation of O$_2$ generated via CO$_2$ dissociation (Tian et al. 2014; Gao et al. 2015; Harman et al. 2015). Second, the extended pre-main-sequence phase (exceeding 100 Myr) of M dwarfs may expose habitable zone planets to high bolometric and extreme ultraviolet (XUV) fluxes, potentially resulting in extensive atmospheric O$_2$ buildup due to H$_2$O photolysis followed by H escape (Luger & Barnes 2015; Wordsworth et al. 2018; Krissansen-Totton et al. 2022).

However, even planets around F, G, and K stars may be susceptible to abiotic oxygen buildup via secular planetary oxidation (Wordsworth & Pierrehumbert 2014; Kleinböhl et al. 2018; Krissansen-Totton et al. 2021). For planets with substantial atmospheric inventories of non-condensing gases (such as N$_2$, CO$_2$, CO), an atmospheric cold trap generally prevents substantial water vapor from reaching the stratosphere (Wordsworth & Pierrehumbert 2014). Consequently, water photolysis and oxygen accumulation following H loss are low. In contrast, small non-condensible background inventories can lead to elevated H$_2$O abundances in the upper atmosphere, producing significant abiotic O$_2$ accumulation (Wordsworth & Pierrehumbert 2014).

This O$_2$ false positive is particularly problematic since it can occur for any stellar type. Indeed, XUV fluxes from more massive stars are large enough (Abrevaya & Thomas 2018) to enable high water photolysis (and H rates) on planets with weak cold traps due to small inventories of non-condensible background gases like N$_2$. Since anticipating atmospheric inventories from first principles is challenging (Stüeken et al. 2016; Wordsworth 2016; Johnson & Goldblatt 2018; Lammer 2019), the atmospheric abundance of N$_2$ must be constrained observationally in order to determine whether $O_2$ in the atmosphere can likely be attributed to remnants of $H_2O$ dissociation and H escape. Constraining the background gas abundance is also crucial for assessing planetary habitability and climate since pressure broadening by $N_2$ (or CO) may significantly modulate planetary radiation budgets (Goldblatt et al. 2009; Paradise et al. 2021).

Unfortunately, $N_2$ has no prominent spectral features within the observational wavelength ranges proposed for a future Habitable Worlds Observatory (HWO), which is traditionally envisioned to have a long-wave observable cutoff of 1.8 μm for exoEarths (The LUVOIR Team 2019; Gaudi et al. 2020). In Earth's atmosphere, $N_2$ does produce $N_2$–$N_2$ collision pairs, $(N_2)_2$, which produce a ∼35% decrease in flux at 4.15 μm (Schwieterman et al. 2015). Additionally, the 4.3 μm $CO_2$ bandwidth corresponds to $(N_2)_2$ absorption for atmospheres with $N_2$ partial pressures above 0.5 bar and significantly widens the band with increasing $N_2$ abundances (Schwieterman et al. 2015), but neither of these near-infrared (NIR) features will be accessible in HWO reflected light spectra. HITRAN identifies weaker $N_2$ absorption features at ∼1.4 and ∼1.09 μm; however, these bands are too weak to be detected by an HWO-type telescope, though they may be more prominent if atmospheric $N_2$ abundances are very large, a possibility we do not consider here.

Without strong spectral features in the nominal HWO wavelength range, one promising possibility is that $N_2$ can be inferred via the absence of spectral features from other plausible background gases that *do* absorb between ∼0.3 and 1.8 μm such as $O_2$, $H_2O$, $O_3$, $CO_2$, etc. By analyzing the directly imaged spectra of an exoplanet, we can infer the background gas indirectly via pressure broadening effect on other gases and by eliminating all other plausible background constituents via a lack of absorption features. This approach has been demonstrated previously. Feng et al. (2018) generated 3D simulated retrievals on a directly imaged Earth analog and successfully constrained both total atmospheric pressure and absorber mixing ratios for $O_2$, $H_2O$, and $O_3$, thereby implicitly obtaining a constrained $N_2$ mixing ratio. Specifically, Feng et al. (2018) were able to simulate a forward model of an Earth analog that produced accurate planetary and atmospheric parameters, implying a background of $N_2$ was constrained, though this was not explicitly quantified in the paper. Similarly, Damiano & Hu (2022) performed simulated reflected light retrievals of terrestrial planets and were able to constrain the main atmospheric component ($N_2$, $O_2$, or $CO_2$) and trace gas mixing ratios ($H_2O$, $O_3$, and $CH_4$) for a variety of atmospheric compositions. Damiano & Hu (2022) reported that optical-only retrievals (0.4–1.0 μm) frequently resulted in incorrect background gas identification, whereas combined optical+NIR retrievals (0.4–1.8 μm) constrained background gases more reliably. Similar techniques have been successfully applied to constrain background gases in simulated retrievals on sub-Neptunes (Damiano & Hu 2021).

However, previous studies have not considered CO backgrounds. This is of particular importance for diagnosing $O_2$ false positives because whereas abundant $N_2$ would disfavor abiotic oxygen accumulation due to low non-condensible inventories, a CO-rich background could indicate abiotic $O_2$ accumulation via $CO_2$ photodissociation (Zahnle et al. 2008; Gao et al. 2015; Hu et al. 2020). Here, we performed simulated reflected light retrievals to investigate whether non-condensible gas abundances could be constrained and whether non-condensible species could be uniquely identified for an Earth analog. We performed simulated reflected light retrievals on directly imaged terrestrial planets using rfast, a new exoplanet atmospheric retrieval suite with direct imaging analysis capabilities (Robinson 2022; Robinson & Salvador 2023). Crucially, we perform retrievals on partial pressures such that all gases are treated equally in our inverse calculations. This ensures that the resultant constraints on background gases are robust and not artifacts of our chosen priors or the initialization of Monte Carlo calculations.

## 2. Methods

For our simulations, we used the rfast retrieval suite, a one- or three-dimensional radiative transfer exoplanetary atmosphere forward model with direct imaging analysis capabilities from Robinson & Salvador (2023), specifically rfast v2.6.5 (Robinson 2022). To generate synthetic spectra of our modeled exoplanet, rfast is equipped with a radiative transfer model that generates a high-resolution spectrum. Molecular opacities in rfast are derived from the HITRAN2020 database (Gordon et al. 2022) using the Line-By-Line ABsorption Coefficients tool, LBLABC (Meadows & Crisp 1996). The spectrum resolution is then degraded to match the resolution specified via an instrument model. Our retrievals were based on reflected light spectroscopy at full phase. To optimize the retrieval process, rfast is nearly entirely written using linear algebra techniques to take advantage of vectorized computational methods, the exceptions being atmospheric recursion relations and integration in the three-dimensional reflected light option. We also made use of rfast's water ice/liquid cloud coverage modeling designed with realistic wavelength-dependent scattering properties and a Henyey–Greenstein phase function. Further notable rfast capabilities include vertically varying gas, temperature, and atmospheric pressure profiles and the ability to divide a synthetic spectrum into multiple bands with distinct wavelength coverage, noise levels, and spectral resolutions. Here we generated our spectra using a constant error function that generated noise at 1 μm for a given signal-to-noise ratio (S/N) and applied that error to each spectral point.

After generating the noisy exoplanet spectra, rfast retrievals are handled via emcee Markov Chain Monte Carlo (MCMC) sampler (Foreman-Mackey et al. 2013). rfast allows for retrievals of atmospheric, planetary, and orbital parameters with uniform or Gaussian priors in either log or linear space. Here, all 14 retrieved planetary parameters were retrieved in log10 space (Table 1). Our retrievals included seven unknown atmospheric gas partial pressures: $N_2$, $O_2$, $H_2O$, $O_3$, $CO_2$, $CH_4$, and CO. A log10 uniform prior is assumed for all constituents ranging from $10^{-2}$–$10^7$ Pa. Note that we are assuming that $H_2$-dominated atmospheres can be excluded based on contextual information, including the fact that combustible $H_2$–$O_2$ atmospheres are kinetically unstable (Grenfell et al. 2018). Yet, our retrievals are agnostic on which of the remaining gases constitutes the dominant background. Our retrievals also include seven unknown planetary parameters: albedo ($A_s$, prior range 0.01–1.0), planet radius ($R_p$, prior range $10^{-0.5}$–$10^{0.5}$ $R_\oplus$), planet mass ($M_p$, prior range 0.1–10 $M_\oplus$), cloud thickness ($\Delta pc$, prior range 1–$10^7$ Pa), cloud top pressure ($p_t$, prior range 1–$10^7$ Pa), cloud extinction optical thickness ($\tau_c$, prior range $10^{-3}$–$10^3$), and cloud cover fraction ($f_c$, prior range $10^{-3}$–1). Cloud top pressures below the surface pressure are prohibited. Each retrieval adopted 200 walkers within emcee and the walkers' positions were randomly initialized within the parameters' prior ranges. Our

Table 1
Nominal Values of Parameters Retrieved

| Parameter Name | Parameter Symbol | True Value (log10) | Prior Range (log10 Uniform) |
|---|---|---|---|
| $N_2$ abundance (Pa) | N2 | 4.896 | [−2, 7] |
| $O_2$ abundance (Pa) | O2 | 4.327 | [−2, 7] |
| $H_2O$ abundance (Pa) | H2O | 2.481 | [−2, 7] |
| $O_3$ abundance (Pa) | O3 | −1.151 | [−2, 7] |
| $CO_2$ abundance (Pa) | CO2 | 1.606 | [−2, 7] |
| CO abundance (Pa) | CO | −1.996 | [−2, 7] |
| $CH_4$ abundance (Pa) | CH4 | −0.695 | [−2, 7] |
| Planetary albedo | As | −1.301 | [−2, 0] |
| Planetary radius ($R_\oplus$) | Rp | 0.0 | [−0.5, 0.5] |
| Planetary mass ($M_\oplus$) | Mp | 0.0 | [−1, 1] |
| Cloud thickness (Pa) | dpc | 4.0 | [0, 7] |
| Cloud top pressure (Pa) | pt | 4.778 | [0, 7] |
| Cloud extinction optical thickness (Pa) | tauc0 | 1.0 | [−3, 3] |
| Cloud cover fraction | fc | −0.301 | [−3, 0] |

**Note.** If not explicitly stated in the figure, reference parameters' values are the ones defined here.

retrievals had each walker take 200,000 steps with a burn-in of 150,000 and a thinning of 100.

Retrieval parameters and their prior ranges, as well as the *true* values used to generate our nominal synthetic spectrum, are shown in Tables 1 and 2. Partial pressure retrievals avoid biased priors compared to retrievals on mixing ratios. Indeed, for the latter case, an a priori background gas must be assumed with a mixing ratio equal to 1 minus the summation of all other constituents. If the other constituents have uniform priors in log space, then the implied prior for the assumed background gas is skewed heavily toward high values. A partial pressure retrieval is broadly equivalent to retrieving centered log ratios for atmospheric abundances (Benneke & Seager 2012; Damiano & Hu 2021, 2022), except that it removes the need to retrieve total pressure as a separate parameter ($p_{max}$ is simply the sum of the partial pressures of every gas in the atmosphere) and avoids any potential biases with MCMC chain initialization in centered-log-ratio space.

Table 2 shows our assumed instrument parameters. Nominal retrievals are performed over the full wavelength range (0.2–1.8 μm), with resolutions of 7, 140, and 70 corresponding to the UV (0.2–0.4 μm), optical (0.4–1.0 μm), and NIR (1.0–1.8 μm) portions of the spectrum, respectively. An S/N of 20 was applied at 1.0 μm with constant error bars and then applied across the full wavelength range. S/N = 20 is assumed in all nominal retrievals, but sensitivity tests for S/N = 10 and S/N = 40 are also performed. An isothermal temperature of 294 K is assumed, and the supplementary material shows that reflected light retrievals are largely insensitive to temperatures within the range expected for an Earth analog.

Using our retrieval outputs from rfast and emcee, we determine what S/N and resolution would be necessary to constrain the non-absorbing atmospheric background's abundance given the HWO wavelength ranges to confidently distinguish $O_2$ false positives. For a more in-depth analysis of rfast's capabilities, see Robinson & Salvador (2023).

Table 2
Nominal Values of the Instrument Model

| Parameter Name | Parameter Symbol | True Value |
|---|---|---|
| Short wavelength cutoff (μm) | lams | 0.2, 0.4, 1.0 |
| Long wavelength cutoff (μm) | laml | 0.4, 1.0, 1.8 |
| Spectral resolving power (lam/dlam) | res | 7, 140, 70 |
| Signal-to-noise ratio | S/N | 20 |
| Wavelength where S/N is Defined | lam0 | 1.0 |

**Note.** If not explicitly stated in the figure, reference instrument model values are the ones defined here and the values shown are in linear space.

We also tested what observation times, telescope diameters, and resolutions would equate to in terms of observational S/Ns at each wavelength. We accomplished this using coronagraph, a reflected light coronagraph noise model for directly imaging exoplanets with a coronagraph-equipped telescope with capabilities for transmission and emission spectroscopy noise modeling. The original IDL code for the coronagraph model was developed and published by Robinson et al. (2016), with the expanded model adapted by Lustig-Yaeger et al. (2019).

## 3. Results

### 3.1. Forward Modeling

As proof of concept, we used rfast to create a synthetic spectrum of our true atmosphere (Table 1) that was then compared to spectra generated with different assumed background gases. Figure 1 shows our nominal (true) $N_2$-dominated spectrum alongside the same planet with different background gases. For the non-$N_2$-dominated cases, the 78% $N_2$ mixing ratio was simply swapped with the alternative background gas. For all plausible alternative background gases ($O_2$, $H_2O$, $CO_2$, CO, $CH_4$, $H_2$), there are prominent absorption features in the 0.2–1.8 μm range, the absence of which (alongside pressure broadening) would strongly suggest either an $N_2$ background gas (or a cosmochemically improbable non-absorbing background such as Ne or Ar—although see the discussion in Section 4 for further consideration of other non-condensibles). Next, we perform simulated retrievals to quantify this result for plausible observational S/N and resolution.

Figure 1 shows there are distinct spectral differences between the spectra of $N_2$-dominated atmospheres with $O_2$, $H_2O$, $H_2$, $CO_2$, and $CH_4$-dominated atmospheres throughout the 0.2–1.8 μm wavelength range. The most notable distinction between $N_2$-dominated and CO-dominated atmospheres occurs at ∼1.6 μm due to a CO absorption feature. $H_2O$ has multiple visible-NIR absorption features, $O_2$ has 0.76 μm absorption and smaller NIR $O_2$–$O_2$ features, $CO_2$ has similar 1.2 and 1.6 μm features, and $CH_4$ has strong absorption throughout the NIR. The opacities in rfast assume $N_2$ broadening since line lists typically do not have data for species other than $N_2$, air, and/or $H_2$. In practice, different background gases would have different broadening effects, so the $N_2$-broadened results here are conservative in that regard.

### 3.2. Simulated Retrievals

Figure 2 shows the retrieved spectrum from our nominal retrieval alongside the true synthetic spectrum with assumed S/N = 20 error bars. The 95% confidence interval on the

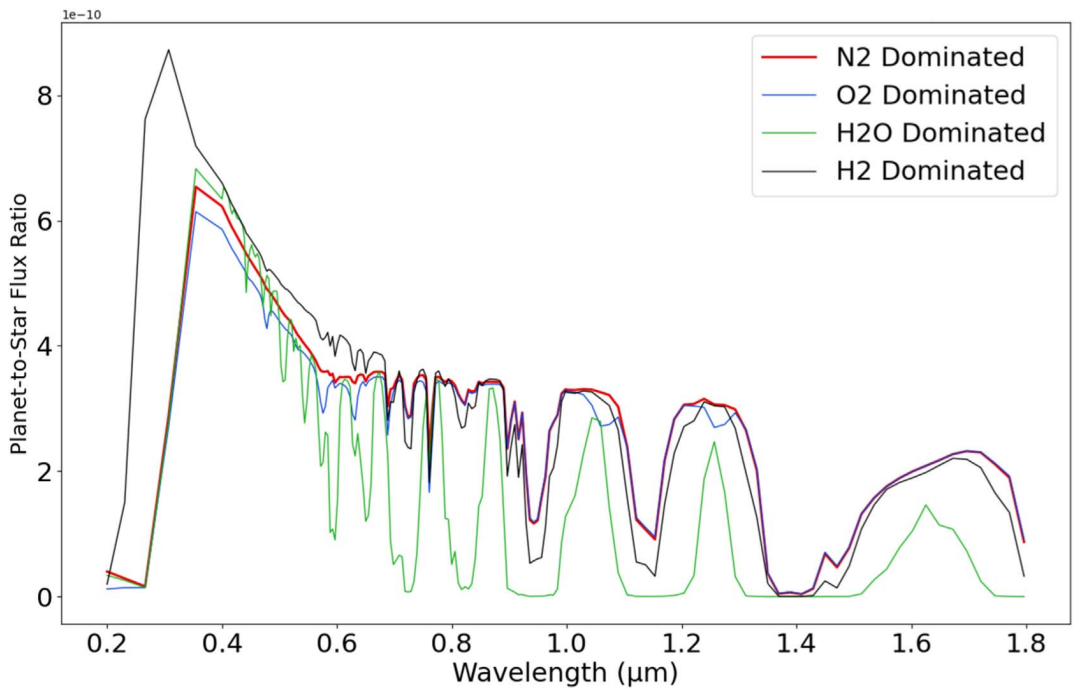

(a) Nominal N$_2$-Dominated Spectrum vs. Non-Carbon Based Atmospheric Gas Dominated Spectra

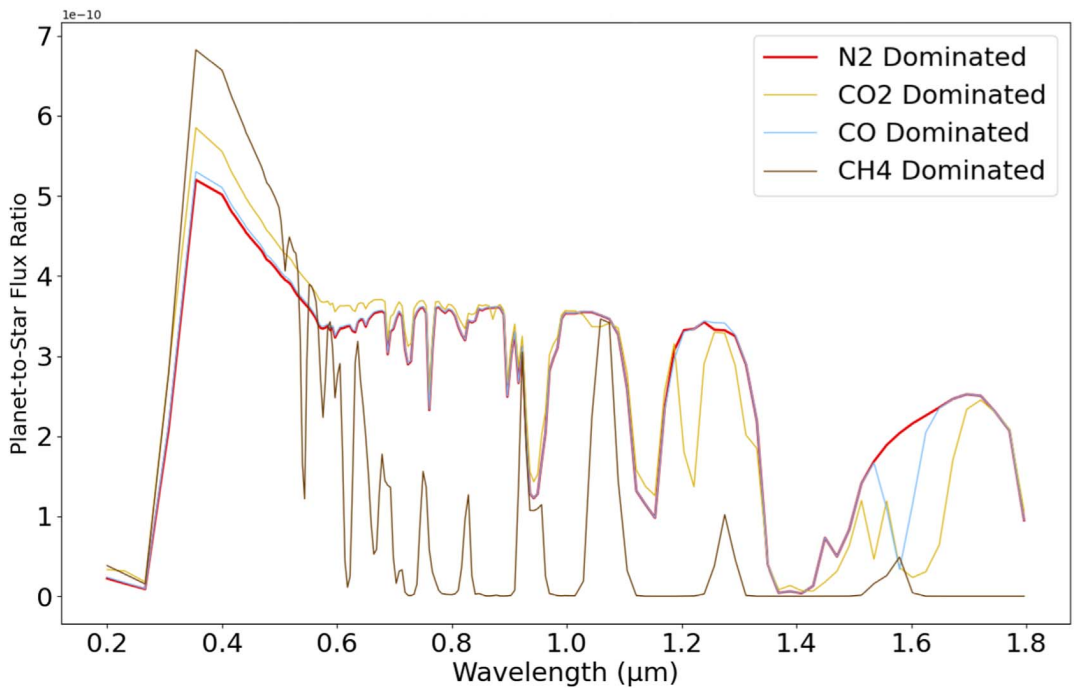

(b) Nominal N$_2$-Dominated Spectrum vs. Carbon Based Atmospheric Gas Dominated Spectra

**Figure 1.** Our nominal N$_2$-dominated model reflectance spectrum compared with spectra generated by swapping atmospheric species abundances with our background N$_2$ abundance (mixing ratio of 0.78). An Earth-like atmospheric composition is assumed, except for the swapped dominant species. (a) Nominal N$_2$-dominated spectrum vs. non-carbon-based atmospheric gas-dominated spectra. (b) Nominal N$_2$-dominated spectrum vs. carbon-based atmospheric gas-dominated spectra.

retrieved spectrum is broadly consistent with the true input spectrum, within error. To generate the confidence interval we sampled 1000 spectra from our posteriors. Additionally, a blue spectrum is overplotted on the confidence interval representing the median (i.e., best-fit) spectrum generated from the spectral confidence interval. Figure 3 shows the corresponding corner plot with posteriors for all retrieved parameters. Every retrieval we ran included the same 14 retrieved parameters, but subsequent corner plots only show gas abundances for easier viewing. Diagonal elements represent the marginalized posterior distributions for each parameter, whereas off-diagonal elements show the 2D covariance between parameters overplotted with the $1\sigma$, $2\sigma$, and $3\sigma$ contours. Along with the parameter name, the 1D parameter posteriors list the median retrieved value with 68% confidence interval uncertainties. Marginalized posteriors also have their parameter's true values (Table 1) overplotted as light blue lines and the dashed lines (left to right) mark the 16%, 50%, and 84% quantiles.

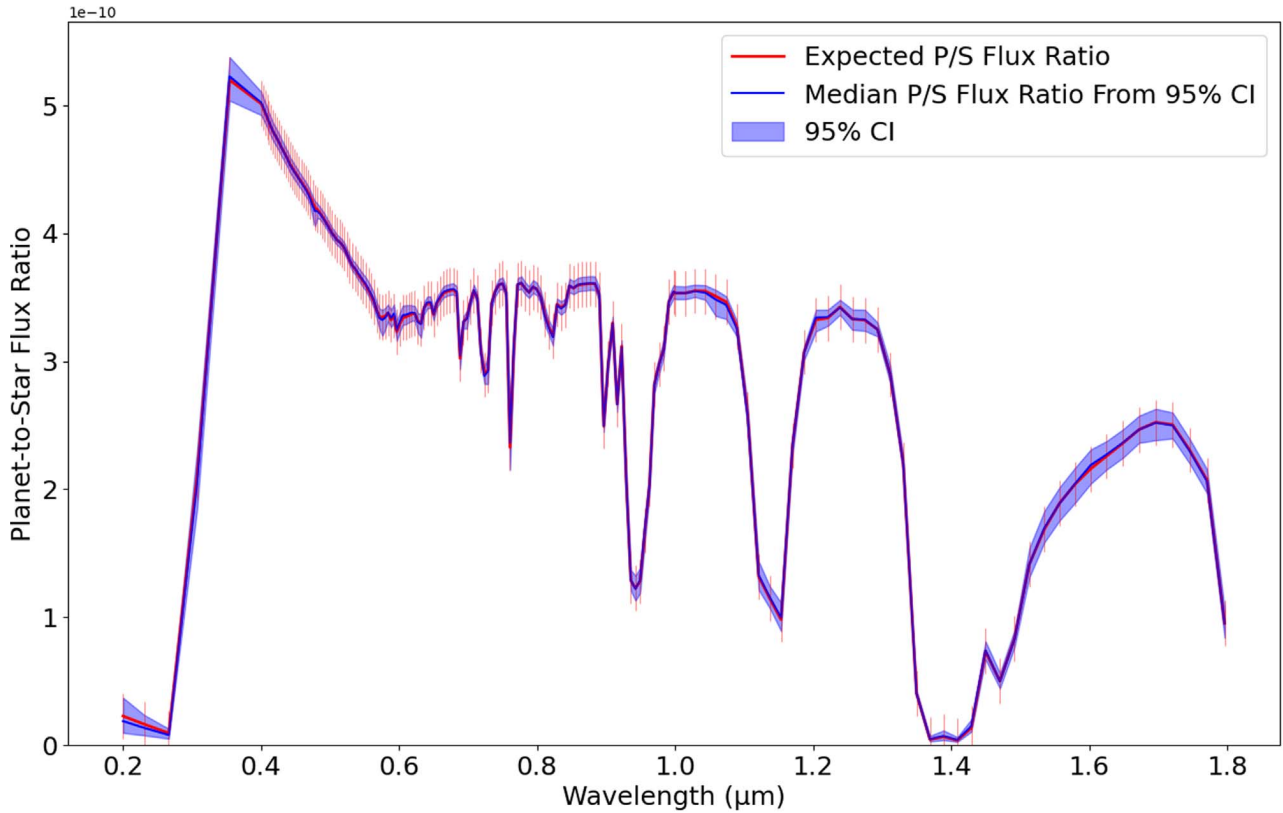

**Figure 2.** True $N_2$-dominated planet-to-star flux ratio spectrum with an assumed S/N = 20 (red) plotted against 1000 randomly sampled retrieved spectra with 95% confidence intervals. The blue spectrum is the median spectrum created from the 1000 plotted model runs that make up our confidence interval.

The posteriors in Figure 3 confirm what was suggested by forward modeling, namely, that it is possible to infer the background gas abundance for direct imaging observations of terrestrial exoplanets. For this Earth-analog case, we infer an $N_2$ background gas with partial pressure $\log_{10}(N_2) = 4.90^{+0.29}_{-0.43}$, corresponding to $N_2 = 7.94^{+7.54}_{-4.99} \times 10^4$ Pa (68% confidence), or equivalently, a mixing ratio of $10^{-0.15^{+0.08}_{-0.27}} = 0.71^{+0.14}_{-0.32}$, shown in Figure 4. The abundances of $O_2$, $H_2O$, and $O_3$ are tightly constrained, which rules them out as background gases, whereas the absence of $CO_2$, $CO$, and $CH_4$ absorption features places upper limits on their respective abundances that similarly rule them out as dominant background gases. Summing the partial pressures of our retrieved atmospheric species we infer total pressure to be $\log_{10}(\mathtt{pmax}) = 5.07^{+0.24}_{-0.22}$, corresponding to $\mathtt{pmax} = 1.17^{+0.87}_{-0.47} \times 10^5$ Pa (with 68% confidence), meaning an $N_2$-dominated background is required to fit the observed spectrum. In other words, there is a high probability that the nitrogen partial pressure is a large fraction of the total atmospheric pressure.

The only possible source of ambiguity is a degeneracy between $O_2$–$N_2$ backgrounds, which can be seen in the $O_2$ posterior and corresponding 2D covariant posteriors of Figure 4. This feature arises because $O_2$ (and $O_2$–$O_2$) absorption features are comparatively weak. While the probability of an $O_2$-dominated atmosphere is low for S/N = 20 retrievals, the $O_2$–$N_2$ background degeneracy can be resolved by higher S/N observations. See Section 3.5 for detailed information regarding how well our model rules out $O_2$-dominated atmospheres for different assumed S/Ns.

Importantly, our atmospheric species were retrieved as partial pressures of the total atmosphere, as opposed to retrieving said species as VMRs. Retrieving the gases as partial pressures allowed our atmospheric constituents' values to be explored independently of each other's abundances and avoid any biases from filling the background with a non-absorbing gas. Although both retrieval methods produced results that were superficially consistent with our true spectra and had confidence intervals that fit within error bars, closer investigation revealed that the VMR retrieval approach was biased toward our predefined background gas (see supplementary materials for a comparison between partial pressure and VMR retrievals).

### 3.3. Simulated Retrievals of CO-dominated Atmospheres

To test whether simulated retrievals can distinguish between $N_2$-dominated and CO-dominated atmospheres, we repeated the calculation described above except that the true $N_2$ and CO mixing ratios were exchanged when generating the true synthetic spectrum. Figure 5 shows the results of this retrieval on a 78% CO atmosphere and Figure 6 compares the retrieved gas abundances for our nominal calculation and CO-dominated retrieval. Figures 6(a) and (b) present the $N_2$- and CO-dominated partial pressure retrieval results, respectively. The CO background is correctly identified in Figure 6(b), with an abundance of $10^{5.03^{+0.24}_{-0.21}}$ Pa, whereas $N_2$ maximum abundance is limited to account for CO absorption at 1.6 $\mu$m. Constraints on other atmospheric constituents are largely unaffected by switching background gases.

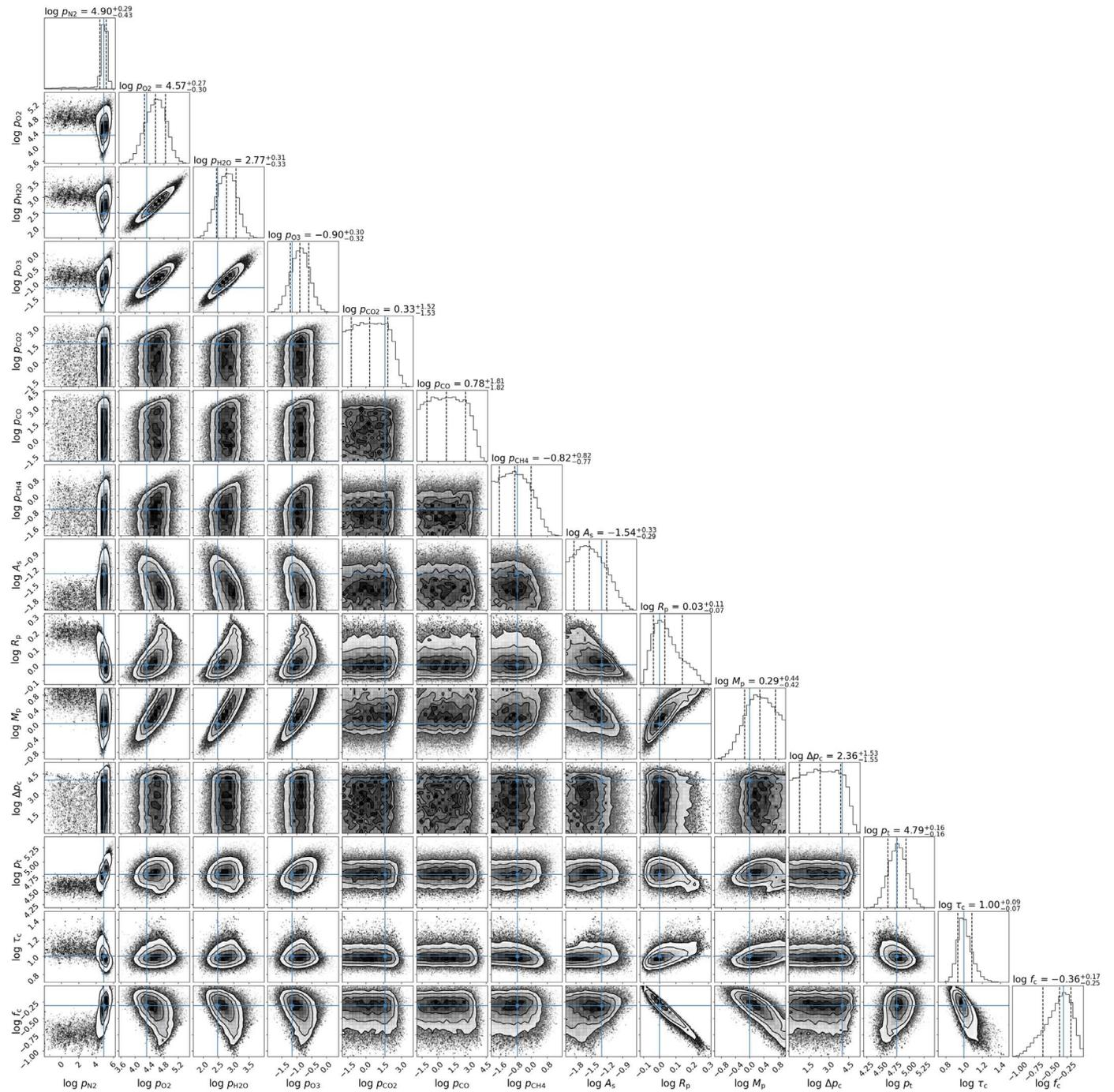

**Figure 3.** Posterior distributions of our nominal partial pressure retrieval with all retrieved parameters. We retrieved on a wavelength-dependent resolution $R = 7, 140, 70$ at wavelength ranges 0.2–0.4, 0.4–1.0, 1.0–1.8 $\mu$m with an assumed S/N of 20 at all wavelengths. The parameters' true values are plotted with light blue lines. The cells off the diagonal are the covariant 2D posterior distributions of their axial posteriors shown with their normal distribution $1\sigma$, $2\sigma$, and $3\sigma$ contours. The diagonal parameter posteriors are also plotted with dashed lines (left to right) that mark the 16%, 50%, and 84% quantiles.

Figure 7 highlights the $N_2$ and CO posteriors for the two partial pressure retrievals described above. Posteriors from the nominal $N_2$-dominated atmosphere retrieval are shown in red, and posteriors from the CO-dominated retrieval are shown in blue. The left column shows the retrieved $N_2$ posteriors for the two retrievals with $N_2$'s true value for each respective retrieval shown with the vertical lines. The right column of Figure 7 illustrates the same for CO abundances, i.e., retrieved CO abundances for both $N_2$- and CO-dominated retrievals. The top row of Figure 7 shows the retrieved gas partial pressures while the bottom row shows the VMRs of the gases. The posterior distributions demonstrate that both $N_2$-dominated and CO-dominated atmospheres can be confidently identified.

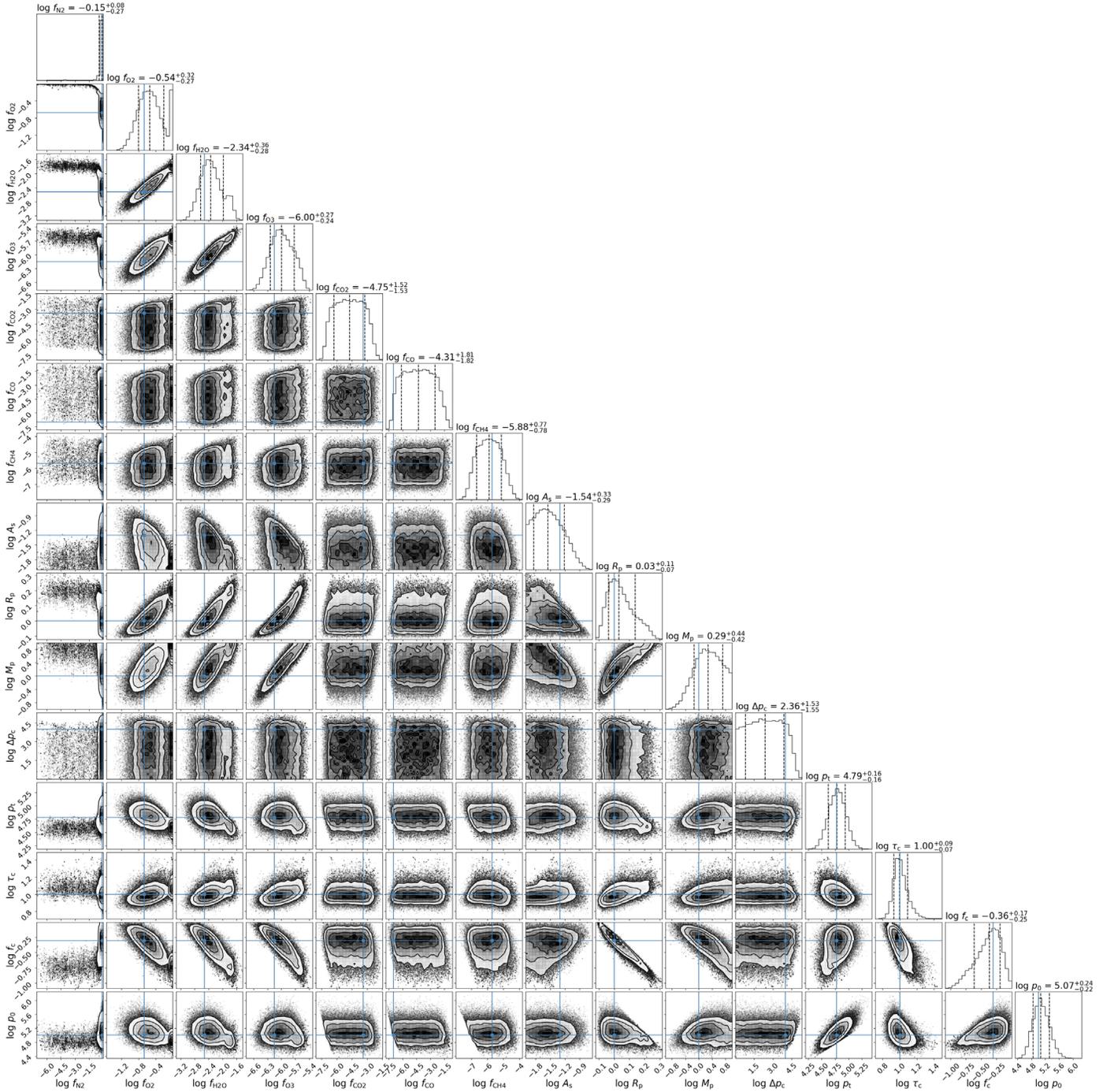

**Figure 4.** The same retrieval and results as Figure 3 except gas values are represented by volume mixing ratios (VMRs) rather than partial pressures. In addition to presenting the atmospheric gases with their VMR values, the surface pressure is also included with respective correlations in the bottom row.

### 3.4. Sensitivity to Observational Wavelength Range

The ability to constrain $N_2$ and CO abundances in the 0.2–1.8 μm wavelength range at S/N = 20 is encouraging. However, we also considered restricted wavelength range observations to test the efficacy of a descoped telescope with a wavelength range of 0.2–1.1 μm. Figure 8 shows our full 0.2–1.8 μm range spectrum with $N_2$–CO abundance tradeoffs, and a highlighted 0.2–1.1 μm range. Crucially, the only spectral feature available to differentiate $N_2$ from CO within the 0.2–1.1 μm band is the Rayleigh tail differences. However, these differences in spectra are imperceptible to a telescope that would have to filter through realistic noise. Note that $N_2$ pressure broadening was assumed for all calculations in this study because CO broadening has not yet been experimentally constrained. In practice, CO versus $N_2$ broadening could help distinguish the two gases, even in the absence of CO absorption features.

All other parameters were held constant for 0.2–1.1 μm retrievals; R = 7, 140, and 70 for wavelength ranges of 0.2–0.4 μm, 0.4–1.0 μm, and 1.0–1.1 μm, respectively, and a wavelength-independent S/N = 20 was assumed. Figure 9 presents the corresponding marginal distributions of $N_2$ and CO for both the 0.2–1.1 μm cases and the 0.2–1.8 μm cases for

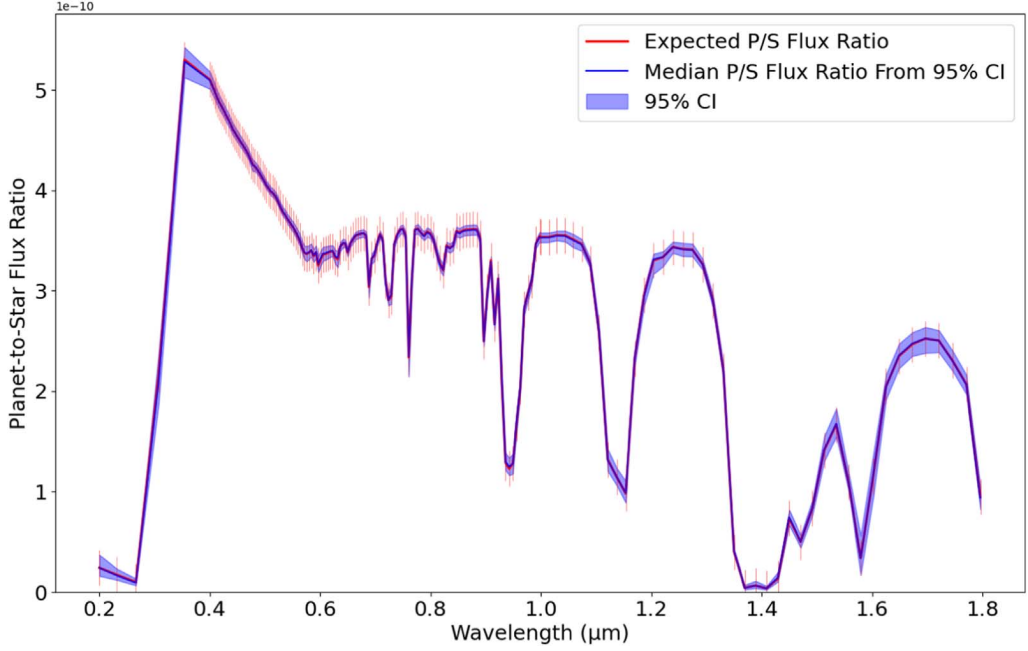

**Figure 5.** True CO-dominated atmosphere planet-to-star flux ratio spectrum (red) plotted against 1000 randomly sampled retrieved spectra with 95% confidence intervals (blue area).

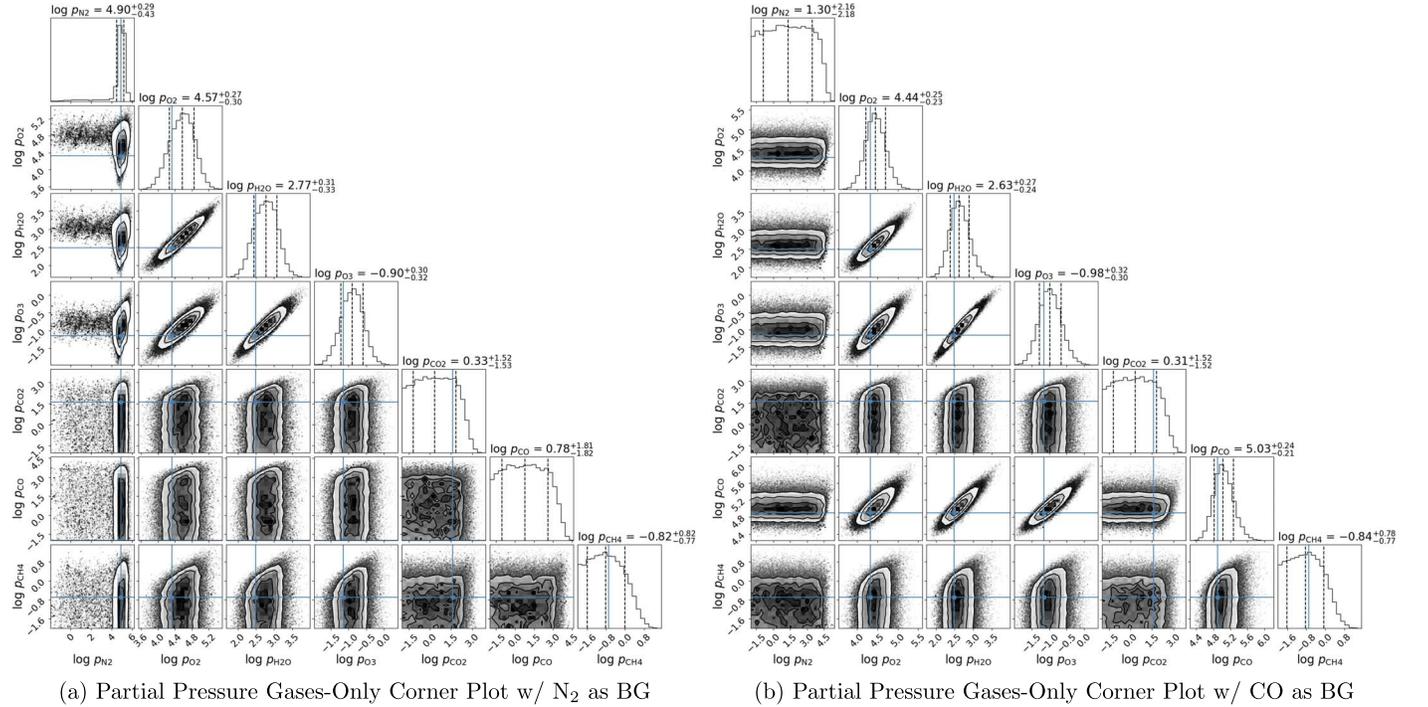

(a) Partial Pressure Gases-Only Corner Plot w/ $N_2$ as BG     (b) Partial Pressure Gases-Only Corner Plot w/ CO as BG

**Figure 6.** Gas-only corner plots (other retrieved parameters not shown) of our nominal $N_2$-dominated partial pressure retrieval and the CO-dominated partial pressure retrieval. Gas abundances are represented by their `log10` pressure in Pa. (a) Partial pressure gas-only corner plot w/$N_2$ as BG. (b) Partial pressure gas-only corner plot w/CO as BG.

comparison and demonstrates our models' abilities to constrain $N_2$ and CO gas abundances in different wavelength regimes. The left column shows our $N_2$ abundance posteriors from our nominal and 0.2–1.1 $\mu$m retrievals, where $N_2$ was defined as the background gas for both cases, so they share the same true value. Similarly, the right column shows our CO abundance posteriors from our nominal CO-dominated and 0.2–1.1 $\mu$m CO-dominated retrievals, again, both cases shared the same true value. As suggested by Figure 9, restricting the wavelength range omits the prominent

1.6 $\mu$m CO absorption feature, and thus $N_2$-dominated and CO-dominated backgrounds cannot be distinguished. For both $N_2$-dominated and CO-dominated true spectra, the $N_2$–CO covariance distribution for the 0.2–1.1 $\mu$m retrieval cases are essentially identical and show that neither CO nor $N_2$ can be confidently and independently detected (see Appendix A.3). Extending observations beyond the 1.6 $\mu$m CO absorption feature greatly tightens constraints on $N_2$ and CO when they are the dominating gas (see the Supplementary Material for $N_2$-dominated and CO-dominated 0.2–1.1 μm confidence intervals).

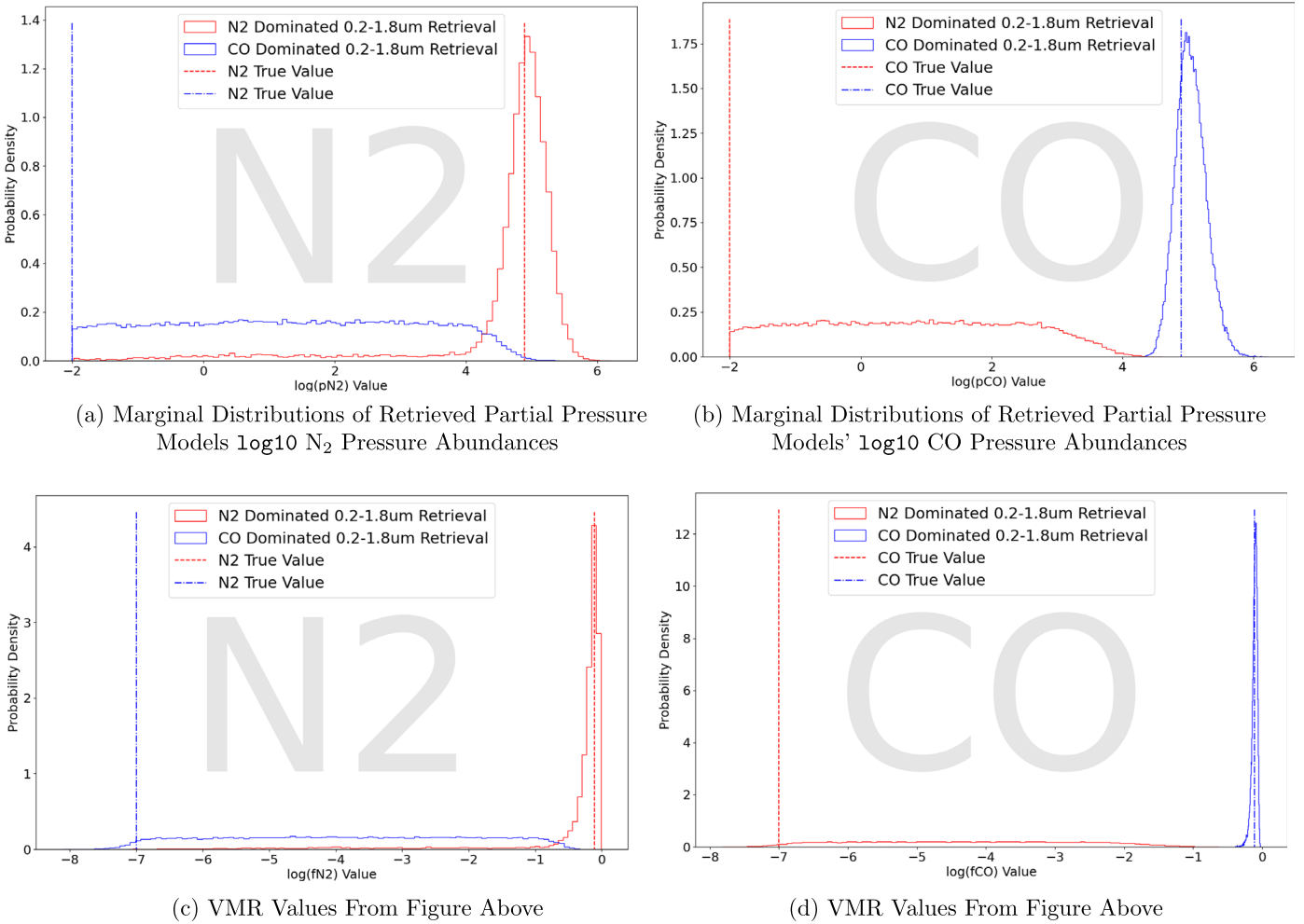

**Figure 7.** Top row (left to right) partial pressure retrieval model $N_2$ and CO posteriors. Bottom row (left to right) VMR adapted partial pressure retrieval model $N_2$ and CO posteriors. Red marginal distributions represent abundances from $N_2$-dominated retrievals and blue distributions represent abundances from CO-dominated retrievals. The gas' true value lines are overplotted for each retrieval illustrated with the same color as its retrieval. The left column presents $N_2$ abundances for both retrievals with CO abundances shown in the right column. Both models are able to constrain our defined background gas and the trace gas has an abundance ceiling that trades with the background gas.

### 3.5. Sensitivity to Assumed S/N

One important consideration for mission design is determining what S/Ns are required for a future telescope to distinguish gas species and their abundances. We ran three retrievals at different representative S/Ns (with otherwise nominal parameter choices) to investigate the minimum S/N required for a telescope to constrain the background gas: S/N = 10, 20, and 40 (Figure 10).

Recall that there was a degeneracy between $O_2$- and $N_2$-dominated atmospheres in the nominal S/N = 20 retrieval presented previously (e.g., Figure 3). This results in a slightly bimodal oxygen VMR marginal posterior (e.g., Figure 4). While the probability ($P$) of an $O_2$-dominated atmosphere is low for S/N = 20 retrievals ($P = 21.86\%$ for $O_2$ VMR > 0.5 and $P = 11.12\%$ for $O_2$ VMR > 0.8), the $O_2$–$N_2$ background degeneracy is resolved at higher S/N observations. At S/N = 40, the probabilities of an $O_2$-dominated atmosphere drop sharply ($P = 0.43\%$ for $O_2$ VMR > 0.5 and $P = 0.0\%$ for $O_2$ VMR > 0.8).

Figure 10 presents the gas-only corner plots of our S/N = 10, 20, and 40 $N_2$-dominated partial pressure retrievals. Figure 10(a) illustrates S/N = 10 retrieval results. $3\sigma$ contours qualitatively differ from the $2\sigma$ distributions of gas values for the S/N = 20 case. Crucially, $N_2$ and CO abundances covary and an $N_2$ background cannot be confidently detected at S/N = 10. Figure 10(b) shows that S/N = 20 results dramatically improve upon the S/N = 10 limitations. Figure 10(c) shows that S/N = 40 provides the tightest constraints on background atmospheric abundances. The low probability degeneracy between an $N_2$ and $O_2$ background is completely resolved at S/N = 40. Retrieved spectra and confidence intervals for the S/N = 10, 20, and 40 cases are shown in the Supplementary Materials.

### 4. Discussion

The ability to constrain background gas abundances is important for ruling out $O_2$ biosignature false positives and for assessing planetary climate. For temperate, Earth-analog planets, $H_2O$ photolysis, and subsequent H escape are typically limited by cold trapping of water vapor in the lower atmosphere. However, small non-condensible (e.g., $N_2$, $CO_2$, CO) background

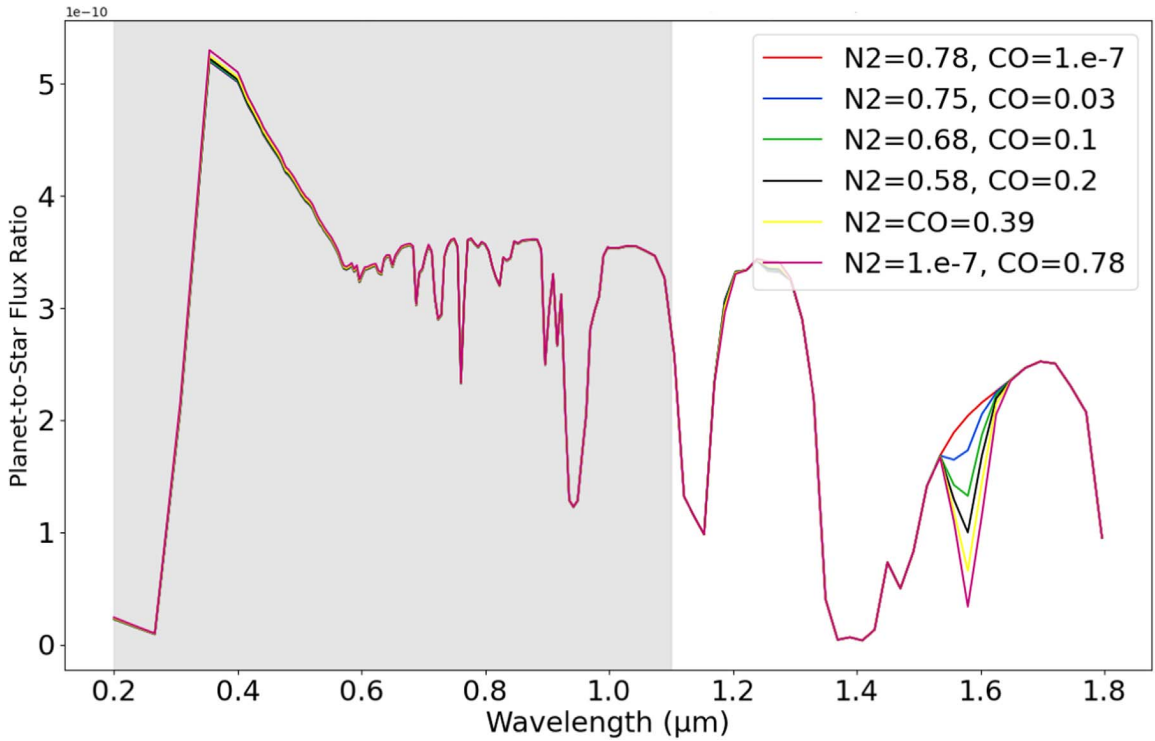

**Figure 8.** Comparison of the nominal $N_2$-dominated atmosphere and trading off VMRs of CO with $N_2$. The $\sim$1.6 $\mu$m CO absorption feature is needed to distinguish between $N_2$ and CO backgrounds.

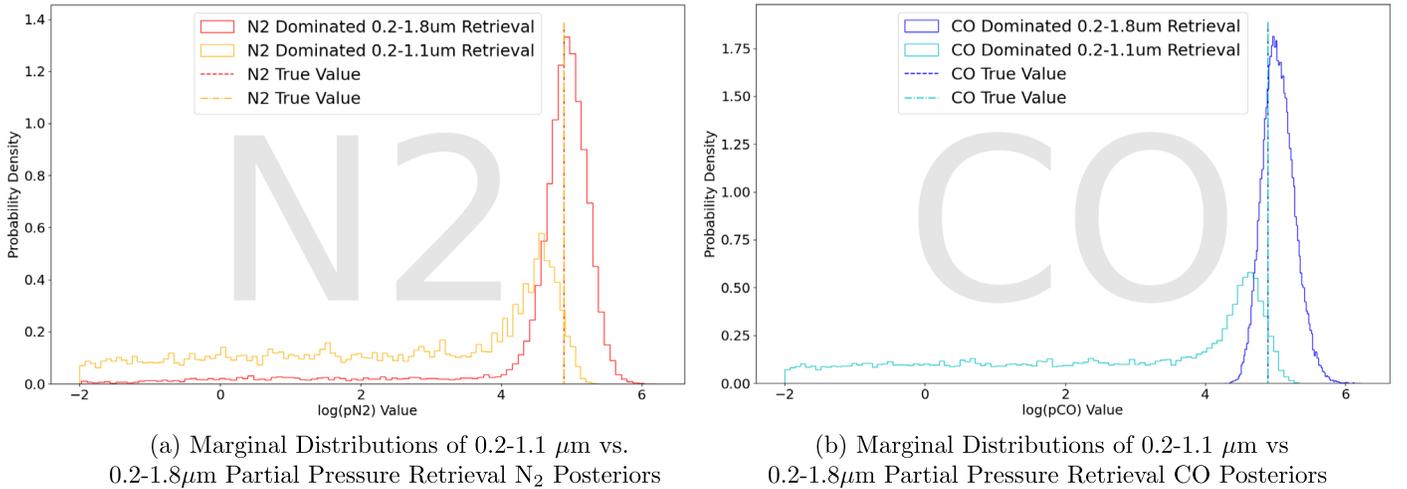

(a) Marginal Distributions of 0.2-1.1 $\mu$m vs. 0.2-1.8$\mu$m Partial Pressure Retrieval $N_2$ Posteriors

(b) Marginal Distributions of 0.2-1.1 $\mu$m vs 0.2-1.8$\mu$m Partial Pressure Retrieval CO Posteriors

**Figure 9.** $N_2$ (left) and CO (right) retrieved partial pressures comparing the 0.2–1.8 $\mu$m to the 0.2–1.1 $\mu$m wavelength ranges. Figure 9(a) presents our $N_2$ abundances for our 0.2–1.8 $\mu$m $N_2$-dominated retrievals in red and our 0.2–1.1 $\mu$m $N_2$-dominated retrievals in yellow. Both $N_2$-dominated retrievals have an $N_2$ abundance true value of 78% of the total atmosphere. Vertical lines show $N_2$'s true value using the same color code. Similarly, Figure 9(b) presents our CO abundances for our 0.2–1.8 $\mu$m CO-dominated retrievals in blue and our 0.2–1.1 $\mu$m CO-dominated retrievals in cyan. Both CO-dominated retrievals have a CO abundance true value of 78% of the total atmosphere. Vertical lines show CO's true value using the same color code. Without extended wavelength coverage, our model is unable to differentiate $N_2$ from CO and they share similar abundances that fall outside of their true value.

inventories can lead to elevated $H_2O$ abundances in the upper atmosphere, resulting in significant abiotic $O_2$ accumulation on geologic timescales (Wordsworth & Pierrehumbert 2014). While the simulated constraints in this study on background $N_2$ and total pressure are crude, they are sufficient to strongly disfavor an extreme H-loss scenario caused by a negligible ($\ll$1 bar) $N_2$ non-condensible background. Moreover, our simulations suggest that $N_2$ and CO backgrounds can be confidently distinguished from S/N = 20 with achievable resolutions $R = 7$, 140, and 70 for wavelength ranges 0.2–0.4 $\mu$m, 0.4–1.0 $\mu$m, and 1.0–1.8 $\mu$m respectively. Even for observations at more restricted wavelength ranges (0.2–1.1 $\mu$m), all other plausible backgrounds ($H_2O$, $CO_2$, $O_2$, $CH_4$) can be ruled out, although the degeneracy between $N_2$- and CO-dominated atmospheres cannot be broken without spectral coverage including CO absorption at 1.6 $\mu$m. This suggests that NIR wavelength capabilities will be crucial for interpreting any $O_2$ detection with next-generation direct imaging observations.

Constraints on total pressure, particularly lower limits, may indirectly provide additional evidence against an oxygen false-positive scenario, even if individual abundances are poorly constrained. This is because the oxygen accumulation scenario envisaged in Wordsworth & Pierrehumbert (2014) is self-limiting—for a pure steam atmosphere, oxygen accumulation from hydrogen escape will itself become the non-condensible background gas, limiting oxygen accumulation at around 0.2 bar. Thus, constraining total pressure to >0.2 bar would add weight to the argument that oxygen is biological. For nominal calculations, retrieved total pressure is inferred to exceed 0.2 bar with 99.99997% confidence. This total pressure constraint is particularly useful for intermediate pressure cases where breaking the $N_2$–$O_2$ degeneracy is more challenging. For example, we conducted a sensitivity test where total pressure = 0.4 bar but mixing ratios were otherwise the same as the nominal retrieval. Here, the $N_2$–$O_2$ degeneracy cannot be broken, even at S/N = 40. However, total pressure is inferred to be >0.2 bar with 96.2% confidence (S/N = 20) and 99.3% confidence (S/N = 40). It should be noted however, that total pressure alone does not unambiguously rule out an oxygen false positive since the precise self-limiting pressure threshold is potentially dependent on planet size, age, other trace constituents, redox evolution, etc.

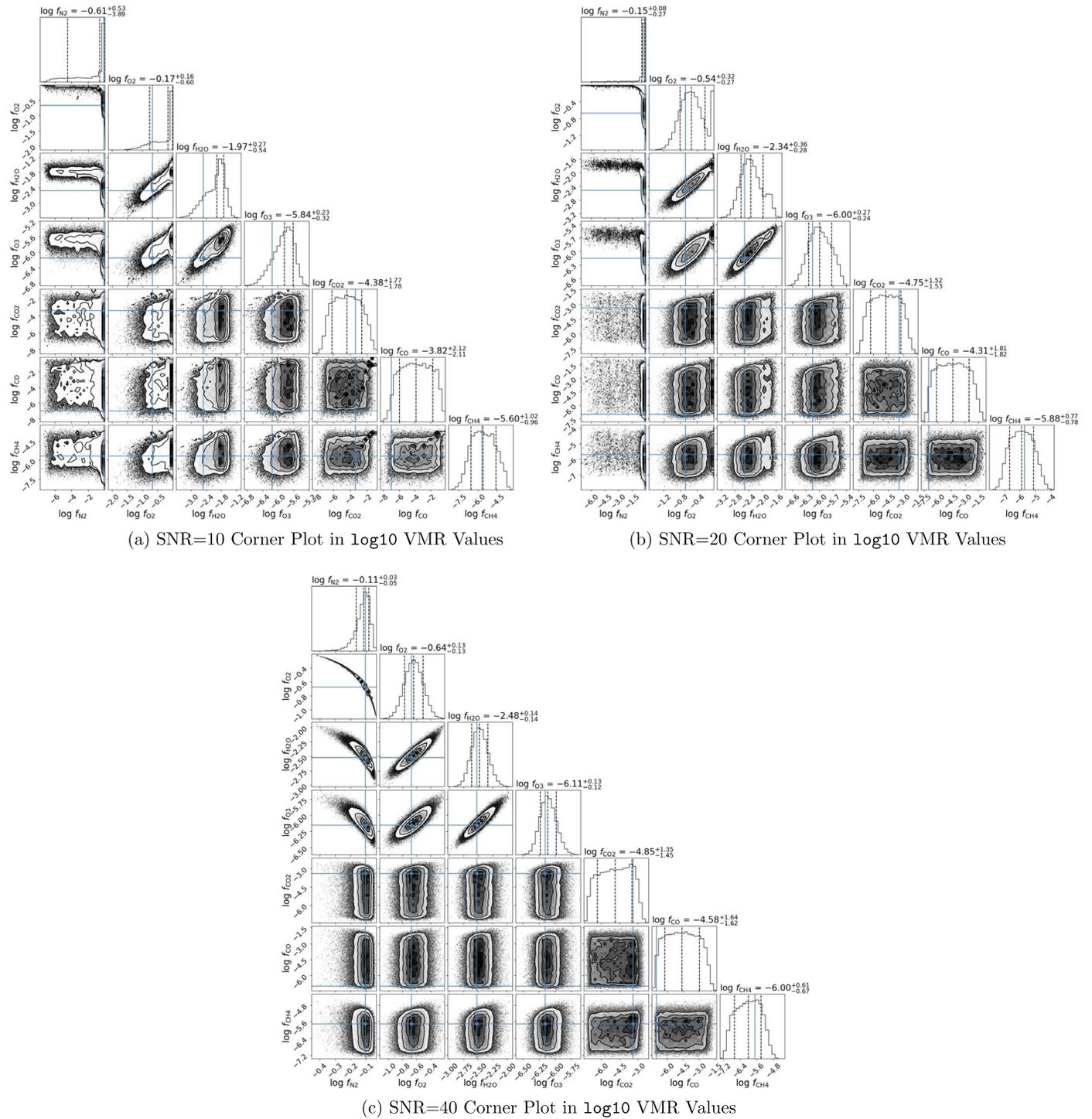

**Figure 10.** Corner plots for our partial pressure retrievals at S/N 10, 20, and 40. All plots are shown with the gas' `log10` VMR values in Pa.

**Table 3**
Hours to Achieve S/N = X at $R = 70$ for a Planet 10 pc Away

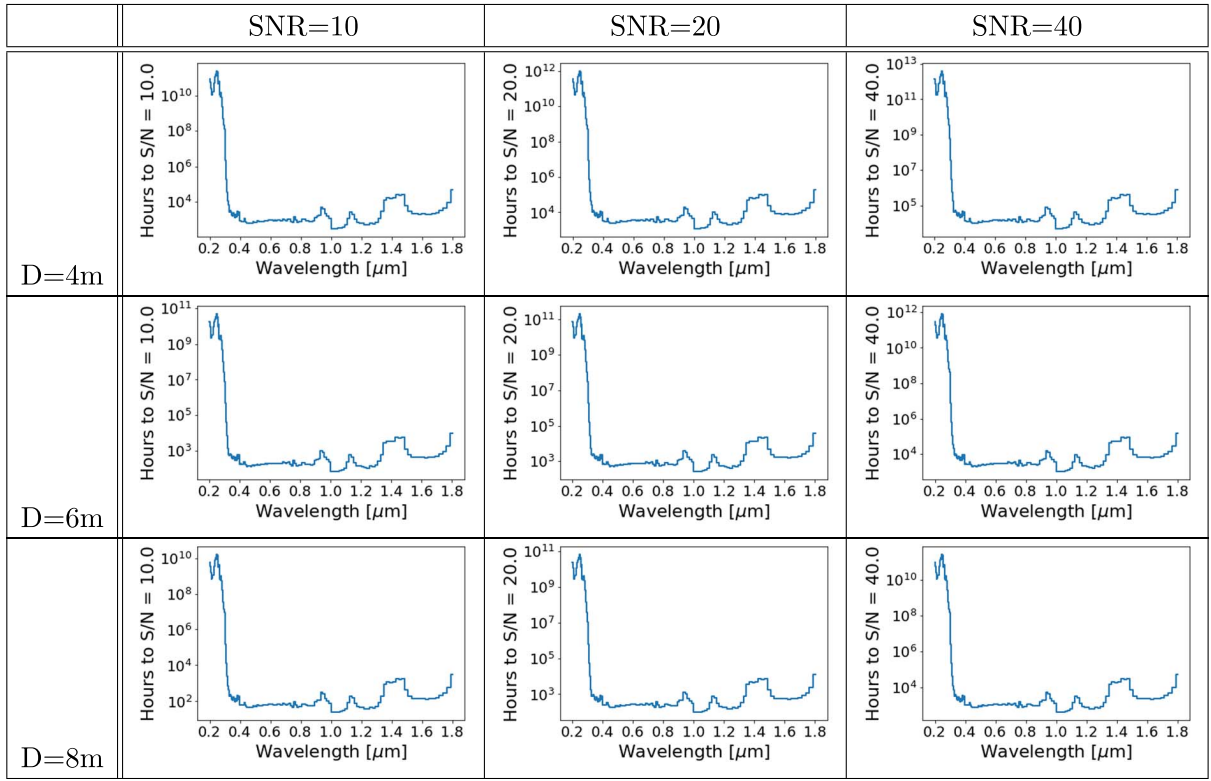

*4.1. Observing Times and Telescope Apertures Requirements*

To illustrate the plausibility of obtaining observations with sufficient S/N and resolution to identify background gases, we used the coronagraph Python package (Lustig-Yaeger et al. 2019) to estimate required integration times for plausible telescope properties. By supplying the model with telescope diameters and desired S/Ns, we calculated the observation duration in hours required to achieve the S/N given for each wavelength point.

Table 3 presents the resulting wavelength-dependent observation times required for an HWO-type telescope with resolution $R = 70$ to achieve specific S/Ns across the 0.2–1.8 $\mu$m wavelength range. The columns are separated into S/Ns of 10, 20, and 40 (left to right), and the rows are separated into telescope diameters of 4, 6, and 8 m (top to bottom). To achieve an S/N = 20 and $R = 70$ at 1.6 $\mu$m for a planet 10 pc away, a 4 m telescope requires ∼8000 hr of observation time, a 6 m telescope requires ∼1600 hr, and an 8 m telescope requires ∼560 hr. The average observation time required to achieve an S/N = 20 and $R = 70$ in the 1.0–1.8 $\mu$m wavelength range for a planet 10 pc away is ∼19,000 hr for 4 m telescopes, ∼4000 hr for 6 m telescopes, and ∼1400 hr for 8 m telescopes. For reference, the Hubble Ultra Deep Field required 550 hr of integration time. Note that the required exposure time can be reduced by integrating spectral information over its wavelength range and will depend on details like detector noise properties and the number of exozodis in the system. The requisite exposure time for gas features that span $N$ spectral elements will (to first order) scale as $1/\sqrt{N}$. For example, if the CO spectral impact spans four spectral elements, there is a factor of ∼2 reduction in exposure time from integrating across the band.

The same process used for Table 3 was also applied to Table 4, except $R = 140$ across the entire 0.2–1.8 $\mu$m wavelength range. This wavelength range is important for identifying $O_2$, $H_2O$, $O_3$, $CO_2$, and $CH_4$ spectral features and constraining their abundances. At $R = 140$ the average observation time required to achieve S/N = 20 within the 0.4–1.0 $\mu$m wavelength range is ∼14,000 hr for a 4 m telescope, ∼2900 hr for a 6 m telescope, and ∼980 hr for an 8 m telescope. While obtaining the longer requisite exposure times at these NIR wavelengths would be challenging, a *deep dive* to try to rule in/out CO would likely be reserved for only key targets. Note also, however, that because the inner working angle (IWA) grows linearly with wavelength for coronagraphs, more distant planets could be lost inside the IWA at these longer NIR wavelengths unless the telescope is larger. For more distant targets, larger apertures would be necessary to enable high enough angular resolution at long wavelengths to see the CO absorption feature at 1.6 $\mu$m.

**Table 4**
Hours to Achieve S/N = X at $R = 140$ for a Planet 10 pc Away

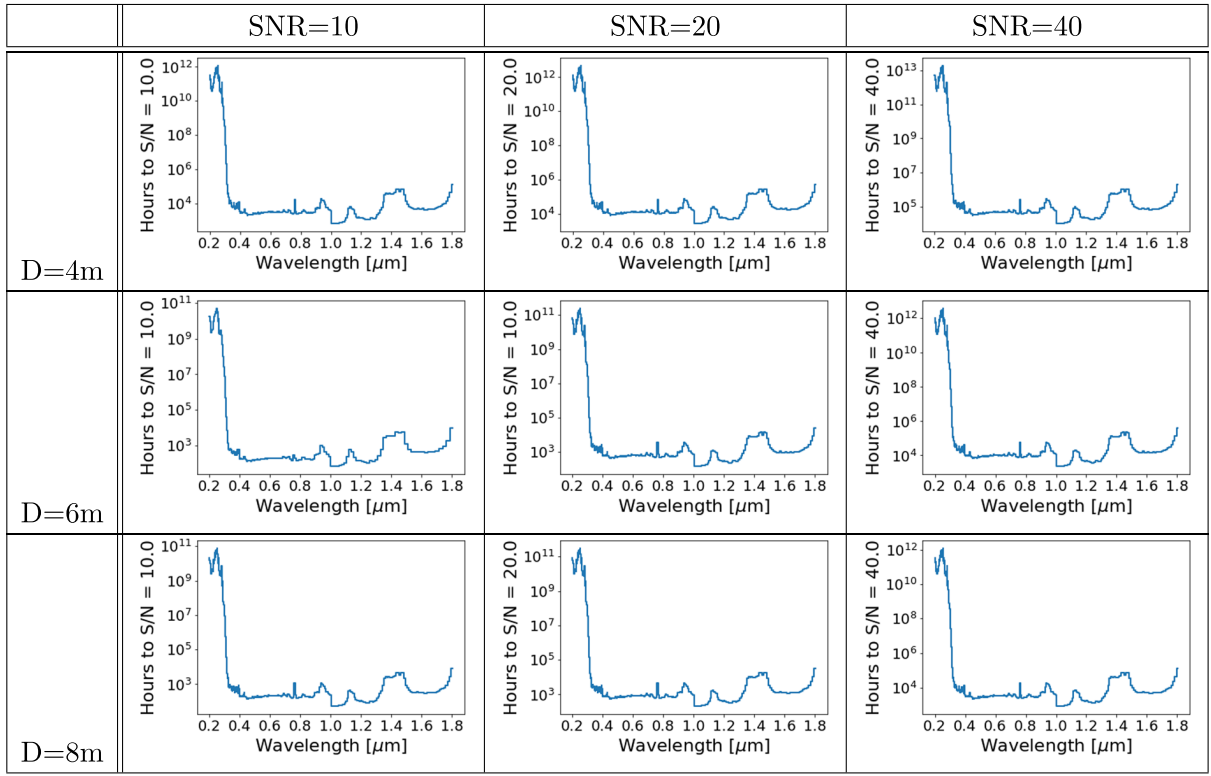

Taken as a whole, these integration time calculations suggest that large apertures (∼8 m) may be necessary to completely break the $N_2$–$O_2$ degeneracy and rule out oxygen biosignature false positives due to low non-condensible gas inventories.

### 4.2. Caveats and Perspectives

Our work in this paper consisted of simulated reflected light retrievals at constant, full phase. Realistic observations would be phase dependent and future work could explore whether observations over different phases could break degeneracies and provide better pressure information to determine background pressure with greater precision via time-variable Rayleigh scattering. In addition, the cloud model consists of water ice and liquid clouds with realistic wavelength-dependent scattering properties and a Henyey–Greenstein phase function. Future work should additionally include wavelength-dependent aerosol absorption for more accurate results.

More broadly, our radiative transfer and retrieval calculations neglect shortwave absorption due to aerosols. This is justified on the grounds that oxygen-rich atmospheres (with water vapor) are unlikely to be hazy due to the oxidation of haze-forming species by OH radicals. The omission of shortwave haze absorption is also unlikely to affect our conclusions regarding disentangling $N_2$ and CO backgrounds since NIR absorption is used to distinguish these gases. However, future work ought to explore the possible influence of hazes on constraining background pressure via the Rayleigh tail, and possible degeneracies introduced by this.

The calculations presented in this paper conservatively assume $N_2$ pressure broadening in all cases. In reality, CO pressure broadening may help distinguish $N_2$- and CO-dominated atmospheres. Future experiments or theoretical work to accurately compute CO pressure broadening could be incorporated into simulated retrievals to determine if differential pressure broadening significantly affects reflected light spectra.

For this work, we have assumed on cosmochemical grounds that Ar, Ne, and other non-condensibles are negligible terrestrial atmospheric background gases. However, the diversity of terrestrial planet atmospheres is unknown, and planetary composition does not scale directly with cosmochemical abundances, nor does atmospheric composition necessarily scale with bulk planet composition. Earth's dry atmosphere contains an Ar mixing ratio of ∼1% (higher than $CO_2$, CO, $CH_4$), resulting primarily from the cumulative radioactive decay of $^{40}K$ over geologic time. For planets with larger inventories of $^{40}K$ and/or older planets with more time for radioactive decay, higher $^{40}Ar$ mixing ratios are conceivable (although the increased heat production and outgassing from enhanced interior heating likely correlates with C, H, and/or N degassing). Scenarios where atmospheric Ar is a significant portion of an ∼0.2 bar atmosphere are conceivable. Future work ought to explore the geochemical plausibility and detectability of these alternative background gases, especially those that are non-absorbing in the UV–visible-NIR.

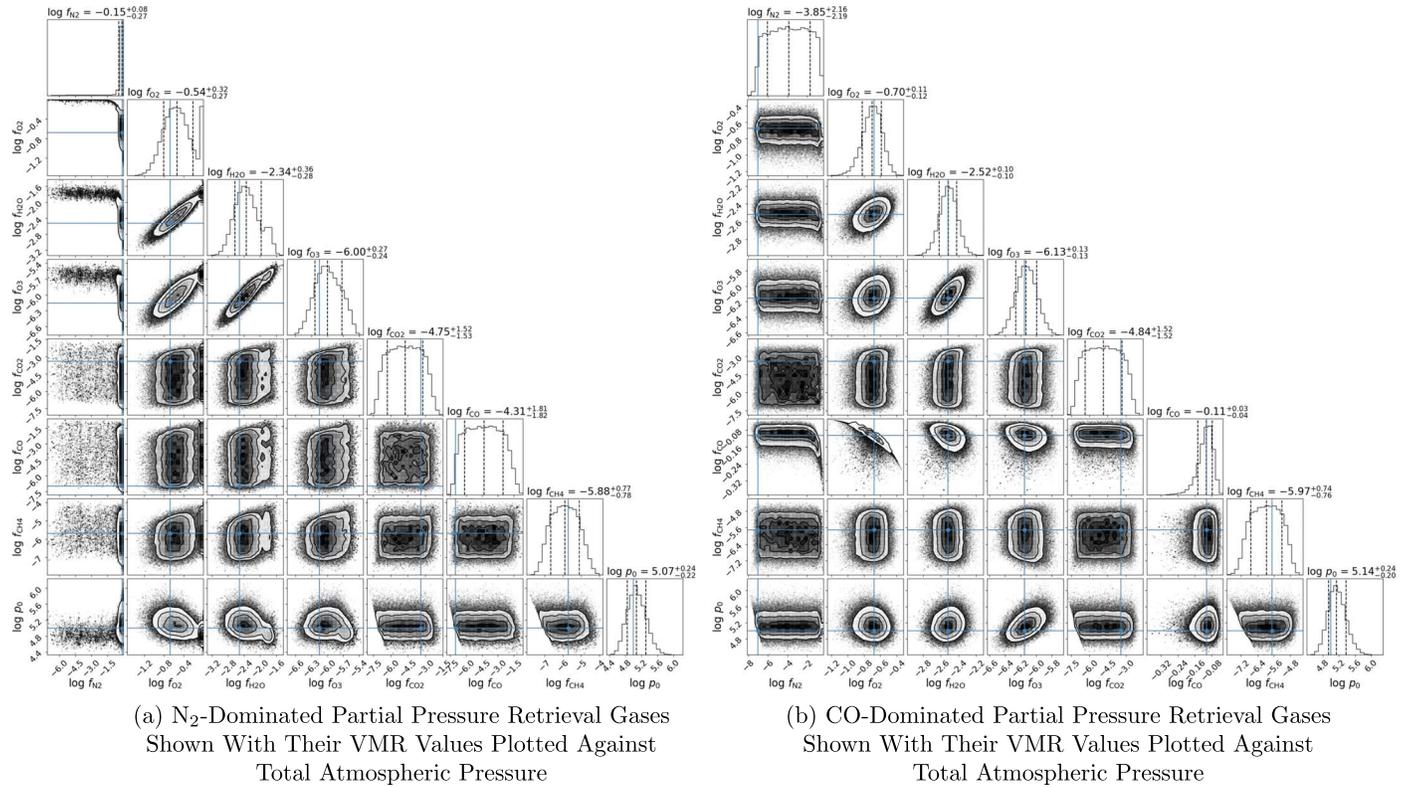

**Figure 11.** Gas-pressure corner plots of $N_2$-dominated (left) and CO-dominated (right) partial pressure retrievals with gas values shown as mixing ratios.

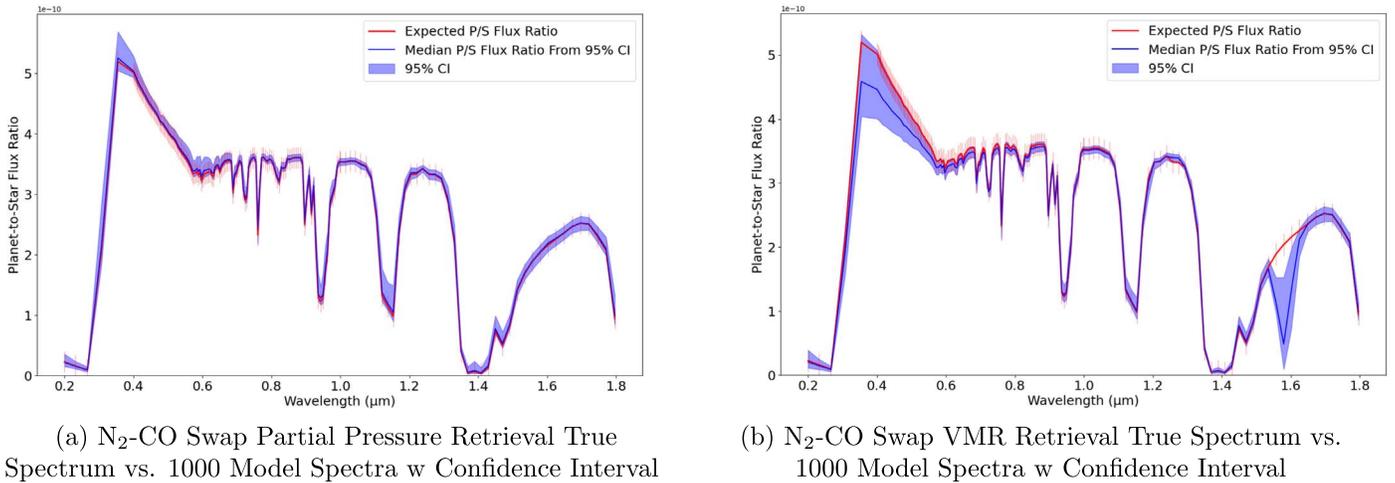

(a) $N_2$-CO Swap Partial Pressure Retrieval True Spectrum vs. 1000 Model Spectra w Confidence Interval

(b) $N_2$-CO Swap VMR Retrieval True Spectrum vs. 1000 Model Spectra w Confidence Interval

**Figure 12.** Spectra confidence intervals for $N_2$–CO swap retrievals. Although both retrievals had true values of $N_2 = 0.78$ and CO = 1.e-7, only the partial pressure retrieval was able to calculate accurate results. We see the $N_2$–CO VMR retrieval sampled many iterations with a high CO abundance evidenced by the absorption feature at ∼1.6 $\mu$m.

The swapped $N_2$–CO abundances retrieval we conducted to explore the $N_2$–CO degeneracy is a somewhat artificial test case. Such a situation is photochemically unlikely because abundant CO was likely sourced from $CO_2$ photolysis, and so a higher background $CO_2$ abundance is to be expected. Conversely, even in the absence of outgassed $CO_2$, CO may be photochemically oxidized to $CO_2$. Thus, another test case we considered is the plausibility of atmospheres with both high CO and $CO_2$ to investigate whether the CO–$N_2$ degeneracy could be broken in the presence of strong $CO_2$ absorption features. Specifically, CO's only significant spectrally active band in the relevant wavelength range (∼1.6 $\mu$m) overlaps with a $CO_2$ absorption feature, and so high levels of CO could possibly be hidden by relatively moderate abundances of $CO_2$, worsening the CO–$N_2$ degeneracy.

To test the plausibility of a HWO breaking CO–$CO_2$(–$N_2$) degeneracies, we ran retrievals using the photochemically self-consistent atmospheres in Gao et al. (2015) at both S/N = 20 and 40. Instrument and atmospheric parameters were otherwise the same as the nominal case except with revised VMRs as follows:

`N2=0.02`

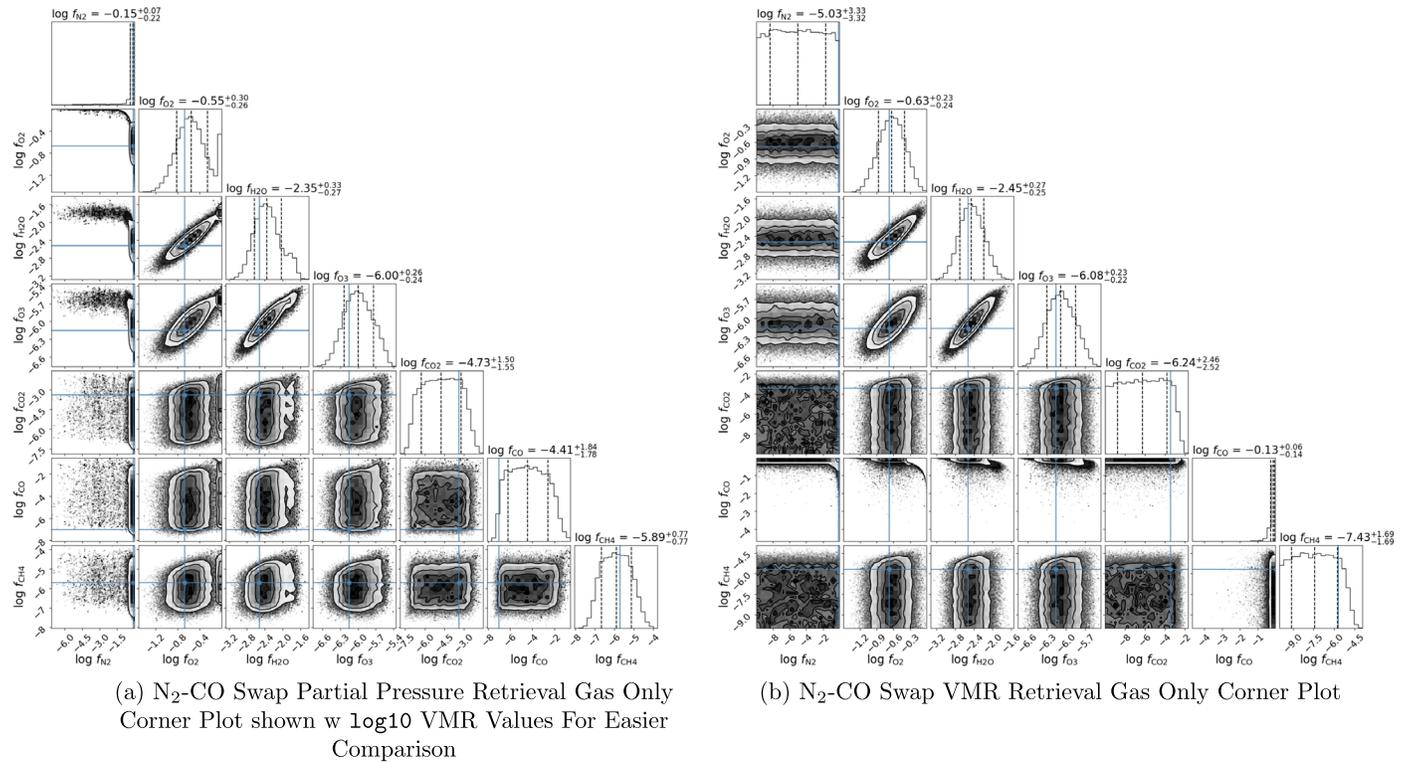

Figure 13. $N_2$–CO swap retrieval corner plots. The VMR retrieval could not constrain either $N_2$ or CO accurately for the swapped case.

$O_2 = 0.17$
$H_2O = 4.e-9$
$O_3 = 7.e-8$
$CO_2 = 0.45$
$CO = 0.35$
$CH_4 = 1.e-7$.

We found that even for these high $CO+CO_2$ atmospheres, a $CO–CO_2$ background can be inferred with high confidence and distinguished from an $N_2$ background (not shown). Specifically, at S/N = 40, a CO mixing ratio of less than ~0.22 is excluded with 84% confidence (and VMR < 0.17 is excluded with 95% confidence) while the $CO_2$ mixing ratio is found to exceed 0.42 with 84% confidence (and $CO_2$ VMR > 0.37 with 95% confidence). Moreover, $O_2$ VMRs can be constrained with high precision even at S/N = 20. Introducing abundant $CO_2$ does not dramatically change our conclusions because the peak $CO_2$ and CO absorption around 1.6 $\mu$m are offset from one another, allowing the two species to be distinguished.

Our work investigated the potential of an HWO to constrain background gas abundances to rule out one type of $O_2$ false positives for Earth analogs following Wordsworth & Pierrehumbert (2014). Future work could convolve simulated $N_2$ partial pressure constraints with atmospheric models of $H_2O$ cold trapping and H escape to determine the extent to which future observations would confidently exclude abiotic oxygen. Additional opportunities for future work include performing retrievals to rule out other $O_2$ biosignature false positives for high $CO_2$ runaway greenhouse atmospheres (Krissansen-Totton et al. 2021), or waterworld false positives (Krissansen-Totton et al. 2021).

## 5. Conclusion

Distinguishing background gases and constraining their abundance can inform assessments of planetary climate and help determine whether $O_2$ in an exoplanet's atmosphere is of biotic or abiotic origin. Our results suggest that distinguishing background $N_2$ from CO and other gases is possible, even for wavelength ranges where $N_2$ has no distinguishable absorption spectral features. This is because all other plausible terrestrial planet background gas constituents absorb in the optical and NIR, and thus can be ruled out as dominant constituents based on the lack of absorption features. Yet, it is important that telescopes like HWO possess wavelength ranges including the CO ~1.6 $\mu$m absorption feature, which is clearly visible at $R = 70$ with an S/N = 20. Additionally, the HWO's observational wavelength range must extend past the CO absorption feature in order to accurately constrain CO abundance, which can then (in combination with total pressure constraints from the Rayleigh slope and gas absorption features) be used to constrain $N_2$ within a wavelength range of 0.2–1.8 $\mu$m. These observations are feasible given plausible integration times. Specifically, an 8 m diameter telescope requires only 560 hr, or ~1/3 the observation time required by a 6 m telescope (1600 hr) with $R = 70$ to achieve a sufficient S/N of 20 at the 1.6 $\mu$m CO absorption feature for an Earth analog at 10 pc.

Bimodal $O_2$ abundance posteriors were obtained for our $N_2$-dominated retrievals, indicating a very low probability of an $O_2$-dominated atmosphere appearing as an $N_2$-dominated atmosphere. However, this $O_2$–$N_2$ is broken at higher S/Ns; for S/N = 20 partial pressure retrievals 21.86% of iterations had an $O_2$ VMR > 0.5 and 11.12% of iterations had an $O_2$ VMR > 0.8, for S/N = 40 retrievals 0.43% of iterations had an $O_2$ VMR > 0.5 and no iterations had an $O_2$ VMR > 0.8.

## Acknowledgments

We thank the UCSC Koret Scholars Program for funding S.H. J.K.T. was supported by the NASA Sagan Fellowship and through the NASA Hubble Fellowship grant HF2-51437 awarded by the Space Telescope Science Institute, which is operated by the Association of Universities for Research in Astronomy, Inc., for NASA, under contract NAS5-26555. This work was additionally supported by NASA Astrophysics Decadal Survey Precursor Science grant 80NSSC23K1471. Finally, we would like to thank the anonymous reviewer, Edward Schwieterman, and the UC Santa Cruz exoplanet squad for their helpful suggestions and critiques. T.D.R. gratefully acknowledges support from NASA's Exoplanets Research Program (No. 80NSSC18K0349) and Exobiology Program (No.80NSSC19K0473), the Nexus for Exoplanet System Science Virtual Planetary Laboratory (No. 80NSSC18K0829), and the Cottrell Scholar Program administered by the Research Corporation for Science Advancement. A.S. and T.D.R. gratefully acknowledge support from NASA's Habitable Worlds Program (No. 80NSSC20K0226).

# Appendix
# Supplemental Material

## A.1. Constraining Total Pressure and Background Gas Composition

By retrieving gas partial pressures, our atmospheric species abundances were independent of each other and of total atmospheric pressure. To calculate our total atmospheric pressure then, we simply summed the retrieved abundances of our atmospheric gases such that

$$\mathtt{pmax} = \mathtt{pN_2} + \mathtt{pO_2} + \mathtt{pH_2O} + \mathtt{pCO_2} + \mathtt{pO_3} + \mathtt{pCO} + \mathtt{pCH_4}.$$

Since total atmospheric pressure is a feature with spectral ramifications (see the Supplemental Material), we are able to constrain our total atmospheric pressure via the summation of our gas abundances.

From our $N_2$-dominated retrievals, we see a strong $N_2$–$O_2$ tradeoff feature in our mixing ratio corner plots. There is additionally a low probability of our model retrieving an $O_2$-dominated atmosphere with small $N_2$ abundances. This $O_2$–$N_2$ degeneracy presents itself in atmospheres lighter than the median pressure expected from an $N_2$-dominated atmosphere. $O_2$ spectral features constrain $O_2$ abundance directly, correlating lower total pressures with low $N_2$ mixing ratios. The degeneracy can be seen in the $\mathtt{O_2}$-$\mathtt{pmax}$ 2D covariant posterior in Figure 11(a). This degeneracy was not seen in CO-dominated atmospheres, which prefer slightly heavier atmospheres, and Figure 11(b) illustrates that concept.

## A.2. Retrieving Partial Pressures versus Retrieving VMRs

Throughout the course of our retrievals, we tested our gas abundance retrievals in two ways: (1) retrieving partial pressure abundances of our gases such that they are independent of each other's values and the total atmospheric pressure is the sum of all constituent partial pressures, and (2) retrieving VMR values of the gases such that they are ratios of the total atmospheric pressure that must sum to 1. The VMR retrieval case coupled the gas abundances with the atmospheric pressure, and since $N_2$ is radiatively inactive in the 0.2–1.8 $\mu$m wavelength range, its abundance was inferred as the difference between unity and the other gas abundances such that

$$\mathtt{fN_2} = 1 - (\mathtt{fO_2} + \mathtt{fH_2O} + \mathtt{fO_3} + \mathtt{fCO_2} + \mathtt{fCO} + \mathtt{fCH_4}).$$

Contrary to our VMR retrieval method, by retrieving partial pressure abundances of our gases we could ensure the retrieval was not biased toward backfilling the atmosphere with any particular gas:

$$\mathtt{pmax} = \mathtt{pN_2} + \mathtt{pO_2} + \mathtt{pH_2O} + \mathtt{pCO_2} + \mathtt{pO_3} + \mathtt{pCO} + \mathtt{pCH_4},$$

where $\mathtt{pN_2}$, $\mathtt{pO_2}$, $\mathtt{pH_2O}$, $\mathtt{pCO_2}$, $\mathtt{pO_3}$, $\mathtt{pCO}$, and $\mathtt{pCH_4}$ values are initialized within their defined prior range (Table 1) then iterated through during the course of the retrieval.

Although both retrieval methods produced results that were superficially consistent with our true spectra and had confidence intervals that fit within error bars, closer investigation revealed that the VMR retrieval approach was biased toward our predefined background gas. We ran VMR and partial pressure retrievals where our forward model was given $N_2$ and CO VMR abundances of $\mathtt{0.78}$ and $\mathtt{1.e-7}$, respectively. Then for our retrieval, we defined CO as our background gas and we swapped the MCMC chain initializations of $N_2$ and CO to see if our models would still be able to find the true value of those gases' abundances.

Figure 12 illustrates the two confidence intervals produced for the $N_2$–CO swap retrievals, with Figure 12(a) displaying results from the partial pressure retrieval and Figure 12(b) displaying results from the VMR retrieval. We can see that for the partial pressure retrieval, our confidence interval follows the true spectrum very well, similar to our nominal $N_2$-dominated partial pressure retrieval. Our VMR $N_2$–CO swap retrieval, however, shows our confidence interval contains many model runs with a high CO abundance, indicated by the 1.6 $\mu$m absorption feature.

Figure 13 displays the gas-only corner plots for our $N_2$–CO swap retrievals. Figure 13(a) presents the partial pressure retrieval, where the blue true value lines are within the 1$\sigma$ contour of the 2D covariant plots and/or the covariant true values converge within the clouds of values. Figure 13(b) presents the VMR retrieval results. Even though $N_2$ had a true VMR of $\mathtt{0.78}$, our model could not constrain it at all, and its value cloud spans the full range it was allowed to explore. Similarly, CO had a true VMR value of $\mathtt{1.e-7}$, but has become our background gas with values centered around $10^{-0.13} = 0.74$. We see for both $N_2$ and CO, their blue true value lines are not found within their 2D covariant plots of values in Figure 13(b).

Thus, we conclude that VMR retrievals should not be used unless the background gas has been confidently identified by previous observations. Backfilling creates a bias in priors that results in incorrect posteriors.

## A.3. Supplementary Plots

Figure 14 shows the true and retrieved spectra for the restricted wavelength range sensitivity test described in Section 3.4. Figure 15 shows the true and retrieved spectra for the S/N = 10, 20, and 40 sensitivity test described in Section 3.5. Figure 16 shows the temperature sensitivity of reflected light spectra is negligible for the temperate planets we are considering. Figure 17 shows the full posterior distribution for retrieved parameters for the CO-dominated sensitivity test described in Section 3.3, and Figure 18 is identical to Figure 17 except that retrieved partial pressures have been converted to mixing ratios.

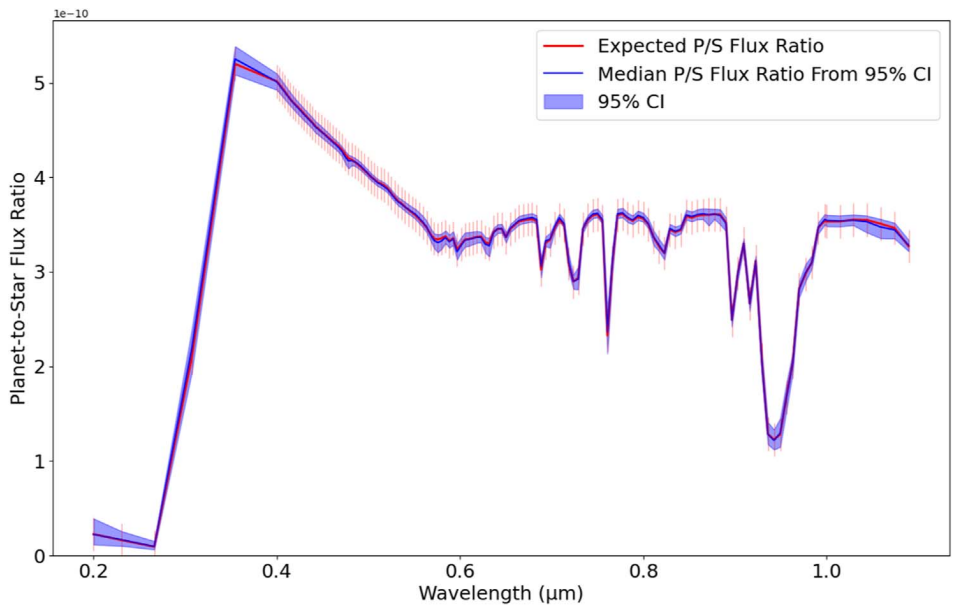

(a) $N_2$-Dominated 0.2-1.1 μm True Spectrum vs
1000 Retrieved Spectra w/ Confidence Interval

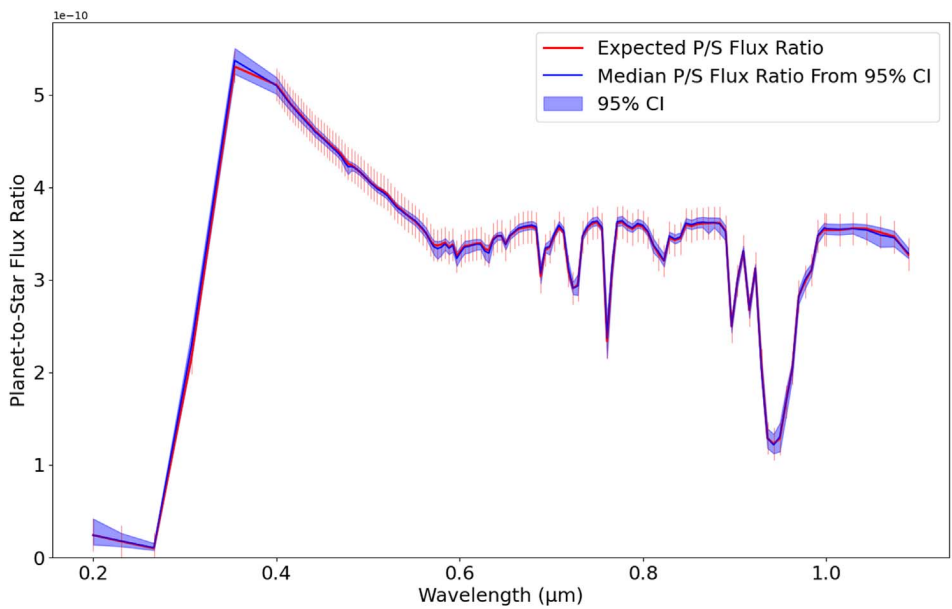

(b) CO-Dominated 0.2-1.1 μm True Spectrum vs
1000 Retrieved Spectra w/ Confidence Interval

**Figure 14.** True 0.2–1.1 μm $N_2$-dominated planet-to-star flux ratio spectrum (top) and CO-dominated spectrum (bottom) plotted against 1000 randomly sampled retrieved spectra with 95% confidence intervals.

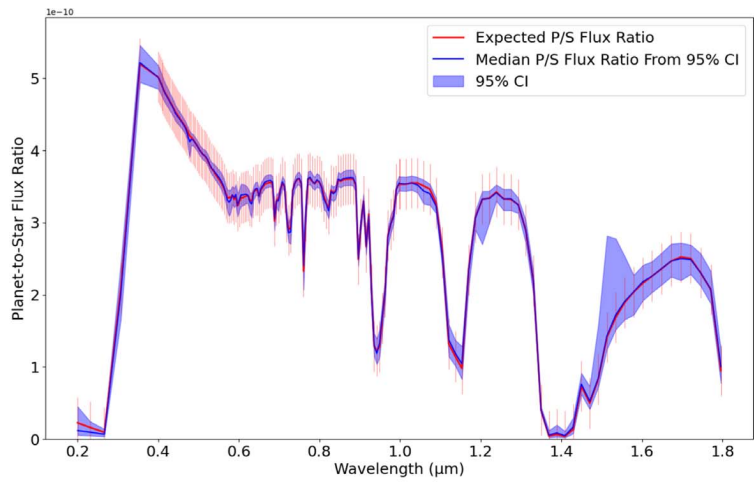

(a) True $N_2$-Dominated Spectrum vs. 1000 Randomly Sampled SNR=10 Retrieved Spectra

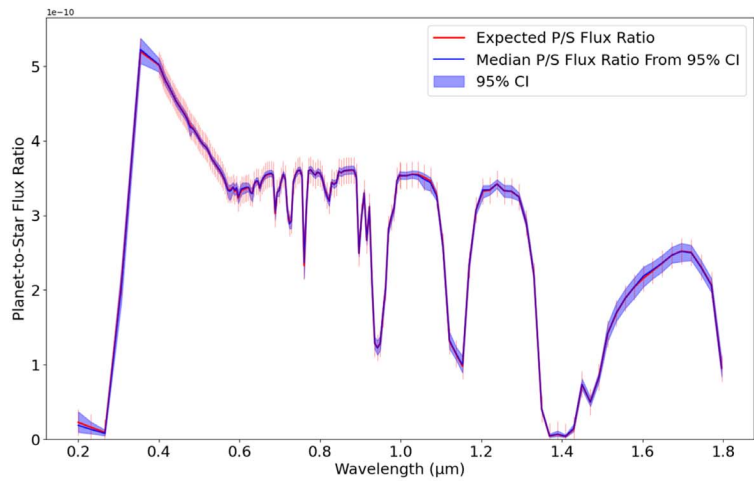

(b) True $N_2$-Dominated Spectra vs. 1000 Randomly Sampled SNR=20 Retrieved Spectra

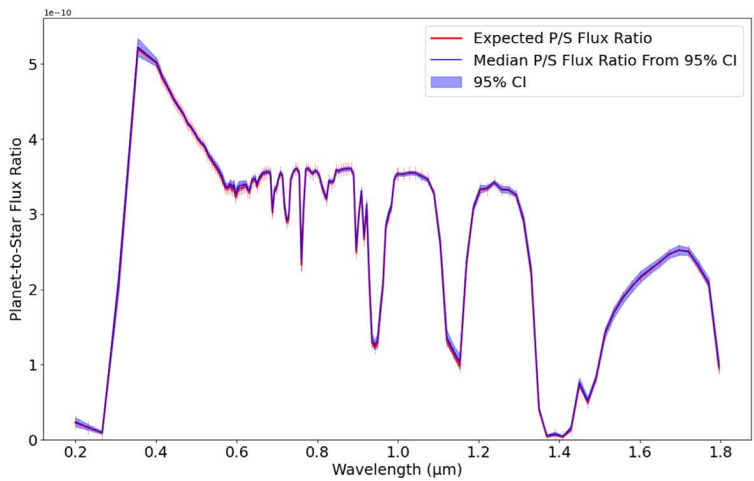

(c) True $N_2$-Dominated Spectra vs. 1000 Randomly Sampled PP SNR=40 Retrieved Spectra

**Figure 15.** $N_2$-dominated partial pressure retrievals' 95% confidence intervals with different S/Ns.

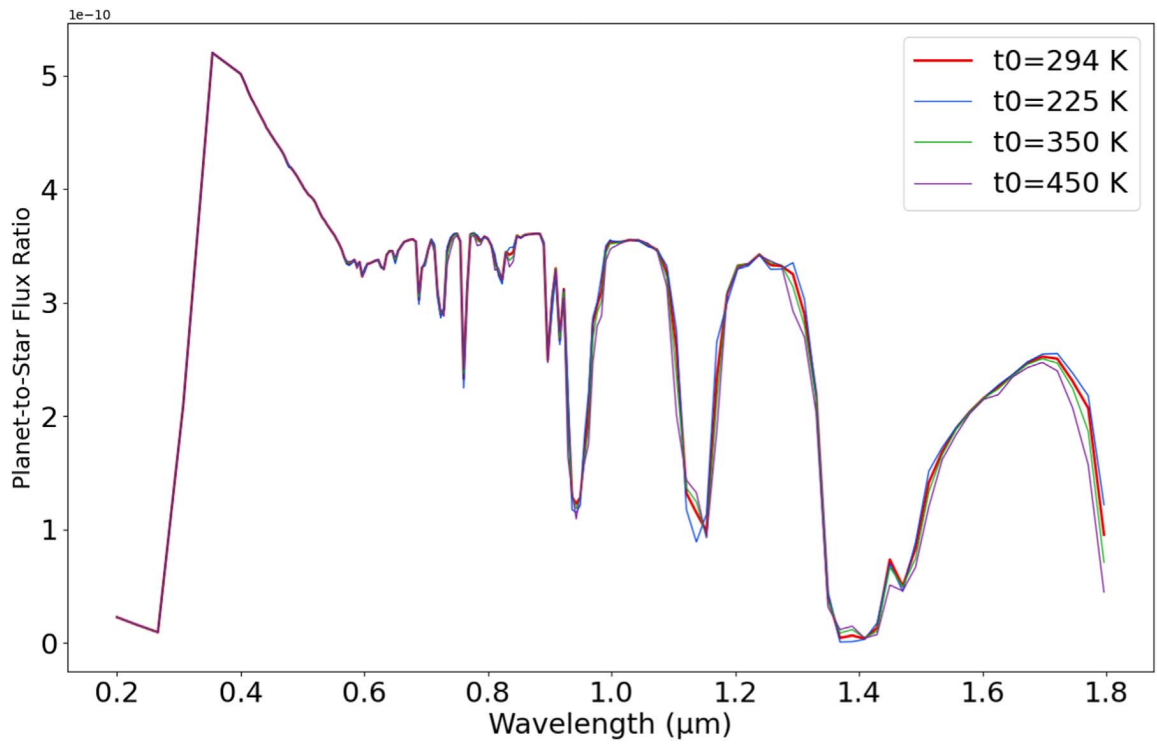

**Figure 16.** Spectra comparison of different surface temperatures on our observed exoplanet. Our nominal model fixes the temperature as outlined in Section 2, `t0 = 294 K`.

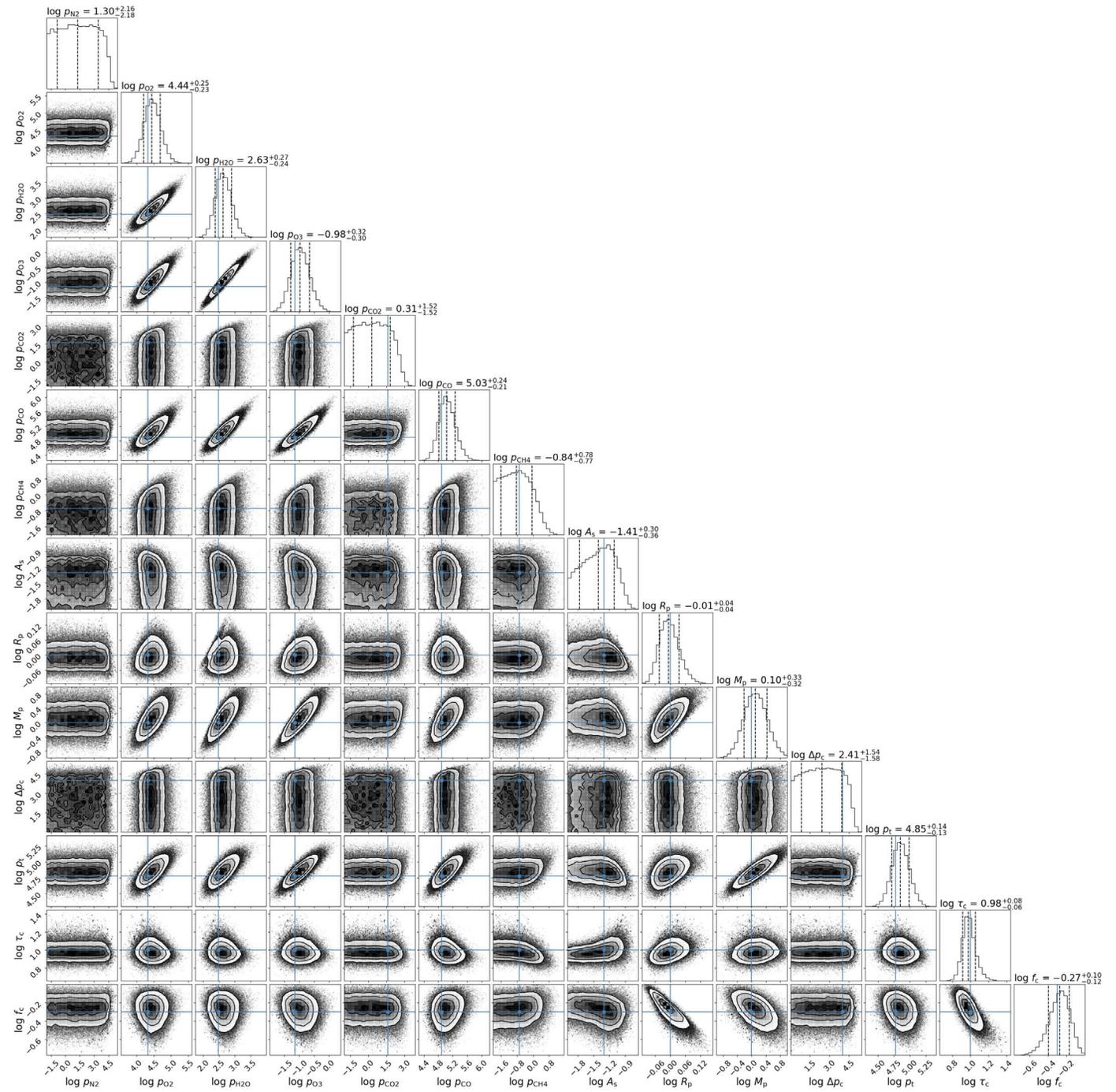

**Figure 17.** Partial pressure CO-dominated retrieval full corner plot.

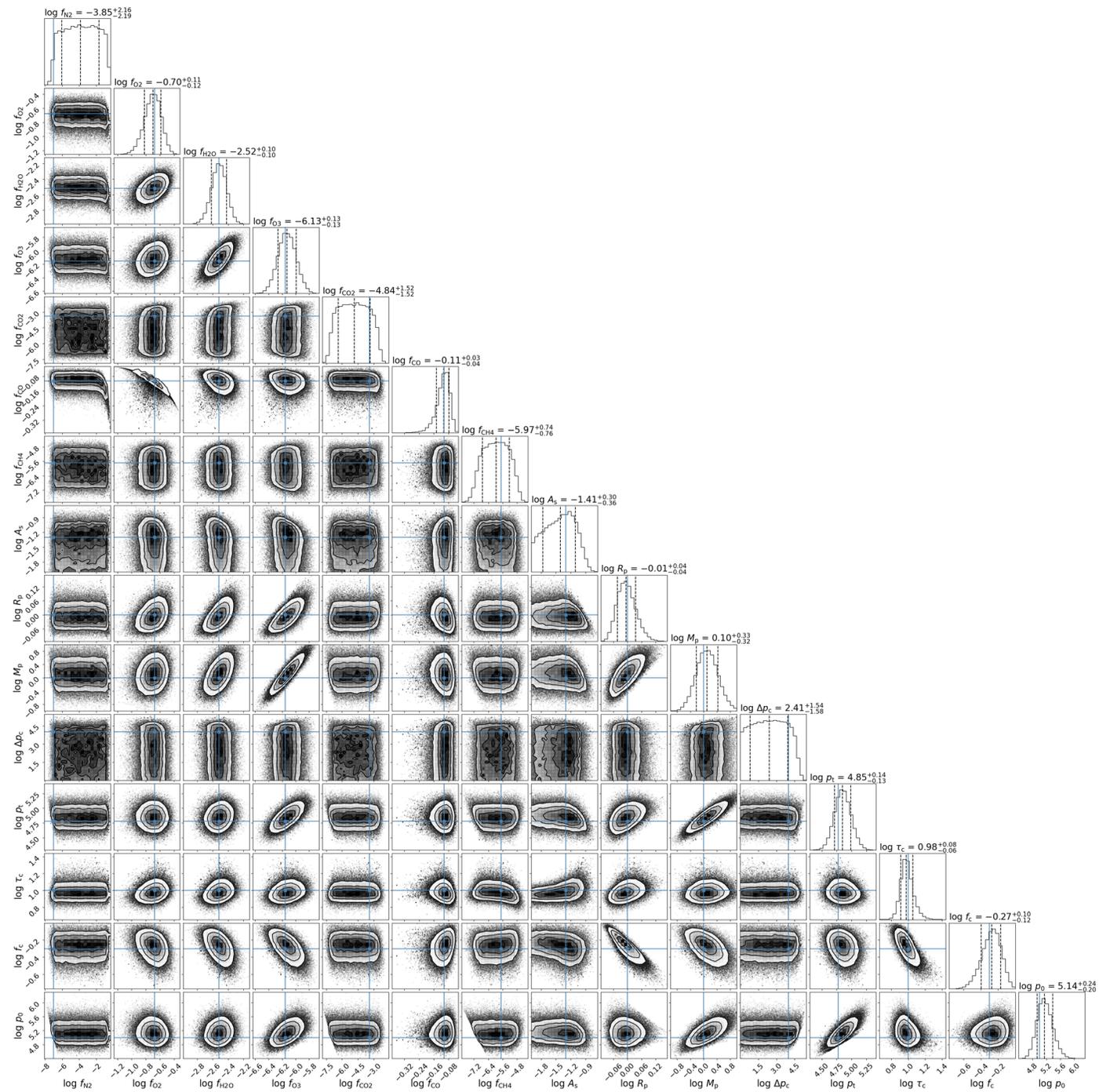

**Figure 18.** Partial pressure CO-dominated retrieval corner plot with gas abundances shown in their VMR values.

## ORCID iDs

Sawyer Hall ⓘ https://orcid.org/0000-0003-2070-5334
Joshua Krissansen-Totton ⓘ https://orcid.org/0000-0001-6878-4866
Tyler Robinson ⓘ https://orcid.org/0000-0002-3196-414X
Arnaud Salvador ⓘ https://orcid.org/0000-0001-8106-6164
Jonathan J. Fortney ⓘ https://orcid.org/0000-0002-9843-4354